\documentclass[a4paper]{article}

\usepackage[english=british]{csquotes}
\usepackage{microtype}
\usepackage{amsmath}
\usepackage{amssymb}
\usepackage{mathtools}
\usepackage{tikz}
\usepackage{xcolor}
\usepackage{graphicx} 
\usepackage{standalone}
\usepackage{listings}
\usepackage{inconsolata} 
\usepackage{booktabs}
\usepackage{pifont}
\usepackage{xspace}

\usepackage{amsthm}
\usepackage{geometry}
\usepackage{hyperref}
\usepackage{cleveref}
\usepackage{subcaption}

\newcommand{\cmark}{\ding{51}}%
\newcommand{\xmark}{\ding{55}}%

\usetikzlibrary{3d,cd,decorations.pathmorphing,fit,perspective,positioning,calc}

\usepackage{menukeys}
\usepackage{fontawesome5}

\usepackage{color} 
\definecolor{codegreen}{rgb}{0,0.6,0}
\definecolor{codegray}{rgb}{0.5,0.5,0.5}
\definecolor{codepurple}{rgb}{0.58,0,0.82}
\definecolor{backcolour}{rgb}{1, 1, 1}

\newcommand\hio{\textsf{homotopy.io}\xspace}

\lstdefinelanguage{OCaml}{
	morekeywords={let, in, if, then, else, match, with, type, exception, open, module, rec, and, fun, function, try, val, do, done, while, not, to, for, of, raise, begin, end, assert, mutable, when, class, object, method, inherit, initializer, new, include, external, constraint, true, false, virtual, lazy, private, public, catch},
	otherkeywords={->, |, :, :=, ;, !, ?, ~},
	sensitive=true,
	morecomment=[l]{(*},
	morecomment=[s]{(*}{*)},
	morestring=[b]",
}

\usepackage[draft,multiuser]{fixme}
\fxsetup{theme=color}
\FXRegisterAuthor{nh}{anh}{NH}
\FXRegisterAuthor{cs}{acs}{CS}
\FXRegisterAuthor{ct}{act}{CT}
\FXRegisterAuthor{jv}{ajv}{JV}
\FXRegisterAuthor{lh}{alh}{LH}
\FXRegisterAuthor{nc}{anc}{NC}

\newcommand{\ignore}[1]{}

\newcommand{\hM}{M^H}

\newcommand\email[1]{\href{mailto:#1}{#1}}

\newtheorem{theorem}{Theorem}
\newtheorem{lemma}[theorem]{Lemma}

\title{\hio: a proof assistant for finitely-presented globular \texorpdfstring{$n$}{n}-categories}

\author{
	Nathan Corbyn
	\thanks{University of Oxford, \email{nathan.corbyn@cs.ox.ac.uk}}
	\and 
	Lukas Heidemann
	\thanks{University of Oxford, \email{lukas.heidemann@cs.ox.ac.uk}}
	\and
	Nick Hu
	\thanks{University of Oxford, \email{nick.hu@cs.ox.ac.uk}}
	\and
	Chiara Sarti
	\thanks{University of Cambridge, \email{chiara.sarti@cl.cam.ac.uk}}
	\and
	Calin Tataru
	\thanks{University of Cambridge, \email{calin.tataru@cl.cam.ac.uk}}
	\and
	Jamie Vicary
	\thanks{University of Cambridge, \email{jamie.vicary@cl.cam.ac.uk}}
}

\begin{document}

\maketitle

\begin{abstract}
	We present the proof assistant \hio for working with finitely-presented semistrict higher categories.
	The tool runs in the browser with a point-and-click interface, allowing direct manipulation of proof objects via a graphical representation.
	We describe the user interface and explain how the tool can be used in practice.
	We also describe the essential subsystems of the tool, including collapse, contraction, expansion, typechecking, and layout, as well as key implementation details including data structure encoding, memoisation, and rendering.
	These technical innovations have been essential for achieving good performance in a resource-constrained setting.
\end{abstract}

\section{Introduction} \label{sec:introduction}

Higher category theory~\cite{leinster2004higher,riehl2022elements} is a branch of mathematics that now has a wide range of applications, in areas as diverse as logic~\cite{hofmann1998groupoid,hottbook}, quantum field theory~\cite{atiyah1988topological,schommer2009classification}, and geometry~\cite{lurie2009higher}.
For working practically with higher categories, string diagrams are an increasingly popular technique, introduced for monoidal categories by Joyal and Street~\cite{joyal1991geometry}, and since extended into higher dimensions by a range of authors~\cite{baez1996higher,barrett2012gray,delpeuch2019complete,dorn2022manifold}.
In a higher-categorical setting, string diagrams take the form of higher-dimensional manifold-like structures, which can efficiently encode complex compositional information.
However, these structures can be hard to visualise, manipulate, or represent in research articles, limiting their effectiveness.

Our tool aims to bridge this gap, allowing string diagrams to become a practical technique for working higher category theorists.
It runs in the browser, giving a low barrier-to-entry, and allows direct construction and manipulation of graphical representations of $n$\-dimensional categorical structures, which we call \emph{$n$\-diagrams}, by direct point-and-click manipulation with the mouse or touch interface.

\begin{figure}
	\captionsetup[subfigure]{justification=centering}
	\begin{subfigure}{0.49\textwidth}
		\centering
		\scalebox{0.75}{\begin{tikzpicture}
\definecolor{generator-1-1-0-pos}{RGB}{192, 57, 43}
\definecolor{generator-2-2-0-pos}{RGB}{243, 156, 18}
\definecolor{generator-3-3-0-pos}{RGB}{142, 68, 173}

\newcommand{\wire}[2]{
  \ifdefined\recolor\draw[color=\recolor, line width=10pt]\else\draw[color=#1, line width=5pt]\fi #2
}
\newcommand{\clipped}[3]{
\begin{scope}
  \newcommand{\recolor}{#1}
  \clip#3;
  #2
\end{scope}
}

\begin{scope}[transparency group]
\fill[generator-1-1-0-pos] (0,0) -- (6,0) -- (6,4) -- (0,4) -- (0,0);
\newcommand{\layer}[1]{
  \clipped{generator-1-1-0-pos}{#1}{(0,0) -- (6,0) -- (6,4) -- (0,4) -- (0,0)}
  #1
}

\wire{generator-2-2-0-pos}{(2,0) -- (2,1)};
\layer{
\wire{generator-2-2-0-pos}{(4,0) -- (4,1) .. controls (4,1.8) and (3.6,2) .. (3,2) .. controls (2.4,2) and (2,2.2) .. (2,3) -- (2,4)(2,1) .. controls (2,1.8) and (2.4,2) .. (3,2) .. controls (3.6,2) and (4,2.2) .. (4,3) -- (4,4)};
}
\end{scope}
\fill[generator-3-3-0-pos] (3,2) circle (0.14);
\end{tikzpicture}}
		\caption{2D projection}
	\end{subfigure}
	\begin{subfigure}{0.49\textwidth}
		\centering
		\includegraphics[width=0.5\textwidth]{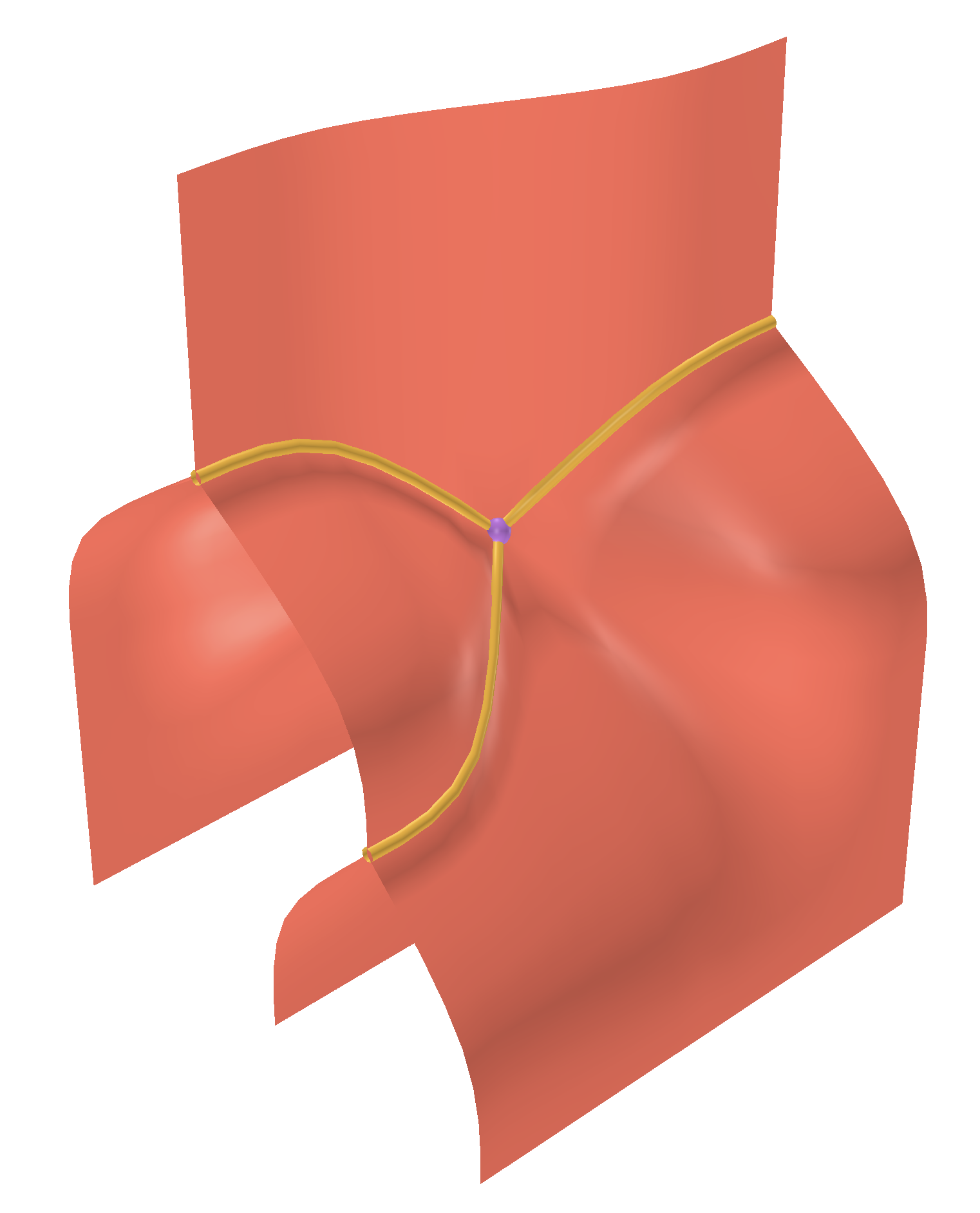}
		\caption{3D projection}
	\end{subfigure}
	\caption{The associator 3-diagram shown in both 2D and 3D (\href{https://beta.homotopy.io/p/2402.00001}{link to online workspace}).}
	\label{fig:associator}
\end{figure}
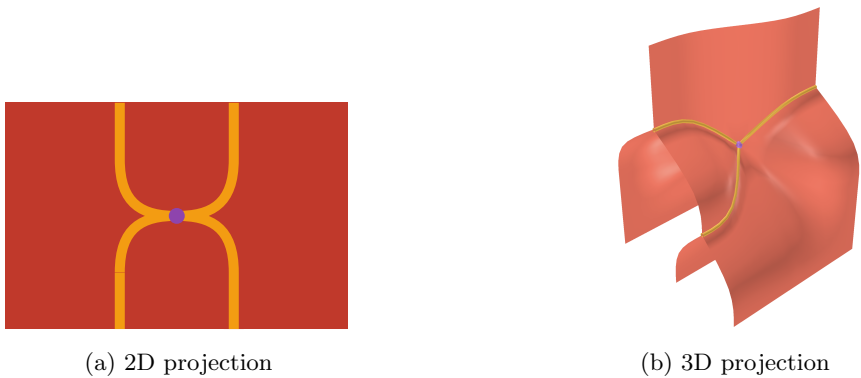

\begin{figure}
	\centering
	\includegraphics[trim={0 15cm 0 8cm},clip,width=0.25\textwidth]{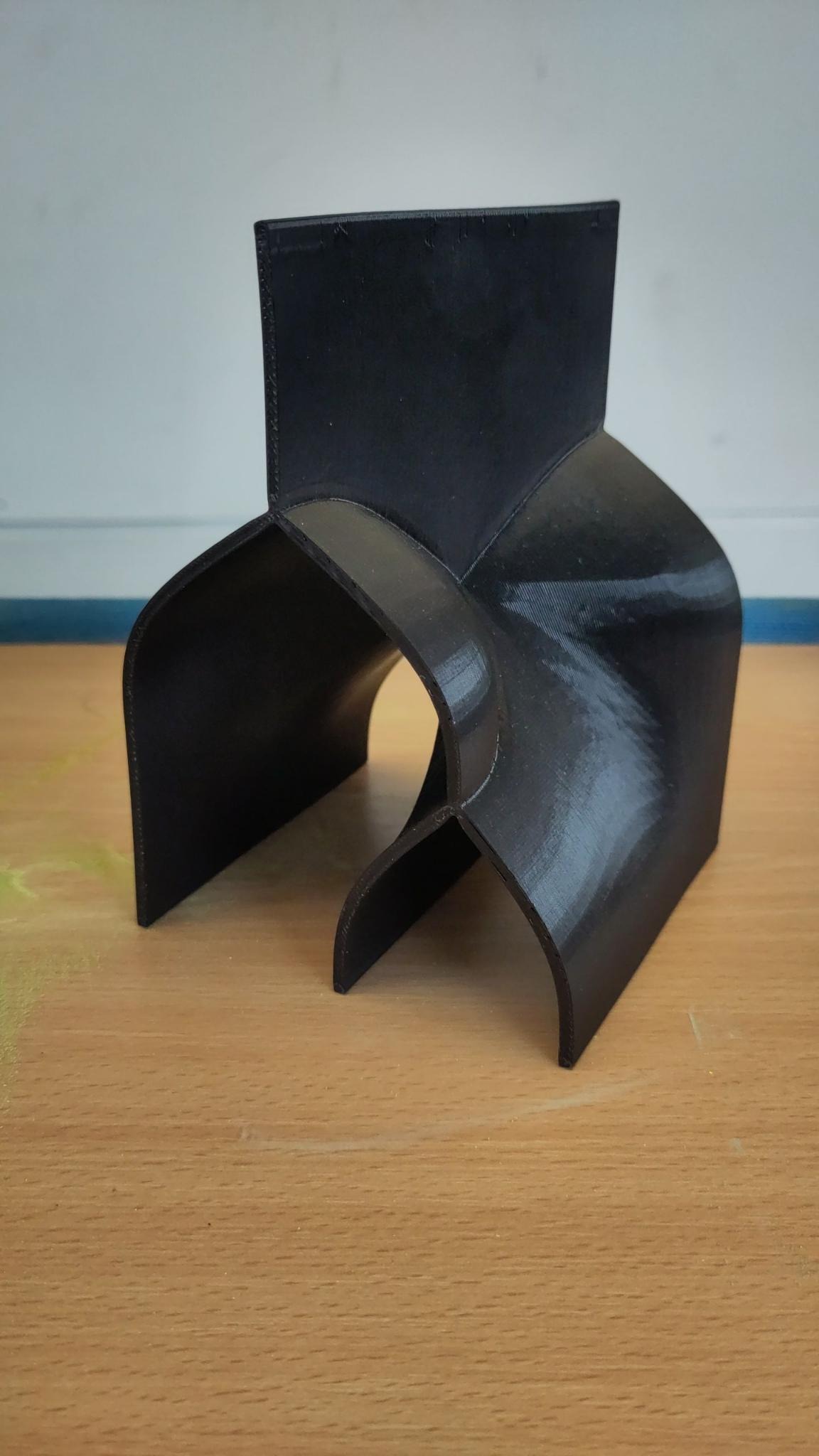}
	\caption{A real-life 3D print of the associator 3-diagram.}
	\label{fig:3dprint}
\end{figure}

In this article, we give an overview of the user interface, and describe the following major subsystems, which give the tool its range of functionality.
\begin{itemize}
	\item \emph{Collapse} acts on a combinatorial $n$\-diagram, constructing a quotient geometry by identifying points which can be considered topologically equivalent.
	\item \emph{Contraction} allows a region of an $n$\-diagram to be geometrically contracted to a point, yielding a homotopy that is itself encoded by an $(n + 1)$\-diagram; the collapse algorithm gives its base case. This is the major mechanism for constructing all nontrivial diagrams in the theory.
	\item \emph{Expansion} provides a limited converse to contraction, defined on a diagram with at least two vertices at the same height, with the effect that one vertex is moved to an adjacent height. The resulting diagram will always contract to yield the original diagram.
	\item \emph{Typechecking} analyses an instance of our $n$\-diagram data structure, and decides whether it represents a valid composite $n$\-morphism in a free higher category.
	\item \emph{Layout} generates a set of linear constraints representing the necessary coordinate relationships between all the parts of a diagram (such as the vertices, wires, and regions), which can be passed to a linear solver, and used by the rendering pipeline.
\end{itemize}
We also examine two significant aspects of the implementation.
\begin{itemize}
	\item \emph{Memoisation} is necessary since the $n$\-diagrams stored by the tool have an intricate recursive structure, which in principle encodes all sub-$k$\-diagrams for $k<n$. If this data was stored separately in memory, the resource requirements of the proof assistant would grow exponentially with diagram dimension. Memoisation ensures each logically-distinct $k$\-diagram is stored only once in memory.
	\item \emph{Rendering} is a complex pipeline that produces suitable output on the screen (see \Cref{fig:associator}); we use SVG for output in dimension 0, 1, and 2, and WebGL for output in dimension 3 and 4. A subdivision algorithm is necessary to produce visually appealing output. We also render to STL for 3D printing (see \Cref{fig:3dprint}), and to TikZ for convenient diagram export.
\end{itemize}

The version of the tool presented in this article is a pre-release version, available here:
\begin{center}
	\url{https://beta.homotopy.io}
\end{center}
It is written in the Rust programming language, and compiled to WebAssembly to run in the web browser.
The implementation is available under a free open-source license on \href{https://github.com/homotopy-io/homotopy-rs}{GitHub}.

\subsection{Mathematical context}

Our proof assistant implements the theory of higher categories known as \emph{associative $n$\-categories}, due to Dorn, Douglas and Vicary~\cite{dorn2018associative,reutter2019high}.
This model is \emph{globular}, in the sense that for $n \geq 2$ the boundary data of any $n$\-cell satisfies the globularity condition: the source of the source equals the source of the target, and the target of the source equals the target of the target.
It is also strictly associative and unital, while retaining weak interchangers; in this sense it is a \emph{semistrict} theory.
It is conjectured that every weak higher category is equivalent to an associative $n$\-category, although the proof of this remains out of reach.

The proof assistant allows users to build composite cells in the graphical language for freely generated semistrict globular $n$\-categories which are free on a \emph{signature}, a list of variables of specified types.
For example, to define a monad-like structure, starting from the empty signature, we might first add an object $x$, followed by a 1-cell $f\colon x \to x$, followed in turn by a multiplication 2-cell $\mu\colon f \circ f \to f$, which we interpret as the monad multiplication.

The tool allows generators of non-zero dimension to be optionally tagged as invertible; this allows the user to work with directed higher categories (if no generators are tagged), higher groupoids (if all generators are tagged), or more general structures.
The resulting invertible structure is coherent, meaning that it automatically satisfies the necessary higher-dimensional constraints, such as the adjunction equations.

\begin{figure}
	\centering
	\begin{tabular}{c c}
\begin{tabular}{@{}c@{}}
\begin{tikzpicture}[xscale=0.6, yscale=0.6]
\definecolor{generator-3-2-0-pos}{RGB}{142, 68, 173}
\definecolor{generator-1-1-0-pos}{RGB}{192, 57, 43}
\definecolor{generator-2-2-0-pos}{RGB}{243, 156, 18}
\definecolor{generator-0-0-0-pos}{RGB}{41, 128, 185}
\begin{scope}
\fill[generator-0-0-0-pos] (0,0) -- (8,0) -- (8,6) -- (0,6) -- (0,0);
\draw[color=generator-1-1-0-pos, line width=2.5pt](2,0) -- (2,3) .. controls (2,3.8) and (2.4,4) .. (3,4) -- (3,6)(5,0) -- (5,2) .. controls (4.4,2) and (4,2.2) .. (4,3) .. controls (4,3.8) and (3.6,4) .. (3,4)(5,2) .. controls (5.6,2) and (6,2.2) .. (6,3) -- (6,6);
\end{scope}
\fill[generator-3-2-0-pos] (5,2) circle (0.14);
\fill[generator-2-2-0-pos] (3,4) circle (0.14);
\end{tikzpicture}
\end{tabular}
\quad & \quad
\begin{tabular}{@{}c@{}}
	\begin{tikzpicture}[xscale=0.9, yscale=0.9]
\fill[red, opacity=.1] (-0.25,-0.25) -- (-0.25,0.25) -- (6.25,0.25) -- (6.25,-0.25);
\fill[green, opacity=.1] (-0.25,0.75) -- (-0.25,1.25) -- (6.25,1.25) -- (6.25,0.75);
\fill[red, opacity=.1] (-0.25,1.75) -- (-0.25,2.25) -- (6.25,2.25) -- (6.25,1.75);
\fill[green, opacity=.1] (-0.25,2.75) -- (-0.25,3.25) -- (6.25,3.25) -- (6.25,2.75);
\fill[red, opacity=.1] (-0.25,3.75) -- (-0.25,4.25) -- (6.25,4.25) -- (6.25,3.75);

\node (R0R0) at (0, 0) {$x$};
\node (R0S0) at (1, 0) {$f$};
\node (R0R1) at (2, 0) {$x$};
\node (R0S1) at (4, 0) {$f$};
\node (R0R2) at (6, 0) {$x$};

\node (S0R0) at (0, 1) {$x$};
\node (S0S0) at (1, 1) {$f$};
\node (S0R1) at (2, 1) {$x$};
\node (S0S1) at (4, 1) {$\beta$};
\node (S0R2) at (6, 1) {$x$};   

\node (R1R0) at (0, 2) {$x$};
\node (R1S0) at (1, 2) {$f$};
\node (R1R1) at (2, 2) {$x$};
\node (R1S1) at (3, 2) {$f$};
\node (R1R2) at (4, 2) {$x$};
\node (R1S2) at (5, 2) {$f$};
\node (R1R3) at (6, 2) {$x$};

\node (S1R0) at (0, 3) {$x$};
\node (S1S0) at (2, 3) {$\alpha$};
\node (S1R1) at (4, 3) {$x$};
\node (S1S1) at (5, 3) {$f$};
\node (S1R2) at (6, 3) {$x$};
\node (R2R0) at (0, 4) {$x$};
\node (R2S0) at (2, 4) {$f$};
\node (R2R1) at (4, 4) {$x$};
\node (R2S1) at (5, 4) {$f$};
\node (R2R2) at (6, 4) {$x$};

\ignore{
{
\node (OR0) at (7, 0) {$R_0$};
\node (OS0) at (7, 1) {$S_0$};
\node (OR1) at (7, 2) {$R_1$};
\node (OS1) at (7, 3) {$S_1$};
\node (OR2) at (7, 4) {$R_2$};
}

{
\draw[->] (OR0) to (OS0);
\draw[->] (OR1) to (OS0);
\draw[->] (OR1) to (OS1));
\draw[->] (OR2) to (OS1);
}
}

{
\draw[->] (R0R0) to (R0S0);
\draw[->] (R0R1) to (R0S0);
\draw[->] (R0R1) to (R0S1);
\draw[->] (R0R2) to (R0S1);

\draw[->] (S0R0) to (S0S0);
\draw[->] (S0R1) to (S0S0);
\draw[->] (S0R1) to (S0S1);
\draw[->] (S0R2) to (S0S1);

\draw[->] (R1R0) to (R1S0);
\draw[->] (R1R1) to (R1S0);
\draw[->] (R1R1) to (R1S1);
\draw[->] (R1R2) to (R1S1);
\draw[->] (R1R2) to (R1S2);
\draw[->] (R1R3) to (R1S2);

\draw[->] (S1R0) to (S1S0);
\draw[->] (S1R1) to (S1S0);
\draw[->] (S1R1) to (S1S1);
\draw[->] (S1R2) to (S1S1);

\draw[->] (R2R0) to (R2S0);
\draw[->] (R2R1) to (R2S0);
\draw[->] (R2R1) to (R2S1);
\draw[->] (R2R2) to (R2S1);

{
\draw[->] (R0R0) to (S0R0);
\draw[->] (R0R1) to (S0R1);
\draw[->] (R0R2) to (S0R2);

\draw[->] (R0S0) to (S0S0);
\draw[->] (R0S1) to (S0S1);

\draw[->] (R1R0) to (S0R0);
\draw[->] (R1R1) to (S0R1);
\draw[->] (R1R3) to (S0R2);

\draw[->] (R1S0) to (S0S0);
\draw[->] (R1S1) to (S0S1);
\draw[->] (R1S2) to (S0S1);

\draw[->] (R1R0) to (S1R0);
\draw[->] (R1R2) to (S1R1);
\draw[->] (R1R3) to (S1R2);

\draw[->] (R1S0) to (S1S0);
\draw[->] (R1S1) to (S1S0);
\draw[->] (R1S2) to (S1S1);

\draw[->] (R2R0) to (S1R0);
\draw[->] (R2R1) to (S1R1);
\draw[->] (R2R2) to (S1R2);

\draw[->] (R2S0) to (S1S0);
\draw[->] (R2S1) to (S1S1);
}
}
\end{tikzpicture}
\end{tabular}
\end{tabular}
	\caption{A string diagram and the corresponding zigzag encoding.}
	\label{fig:stringdiagzigzag}
\end{figure}
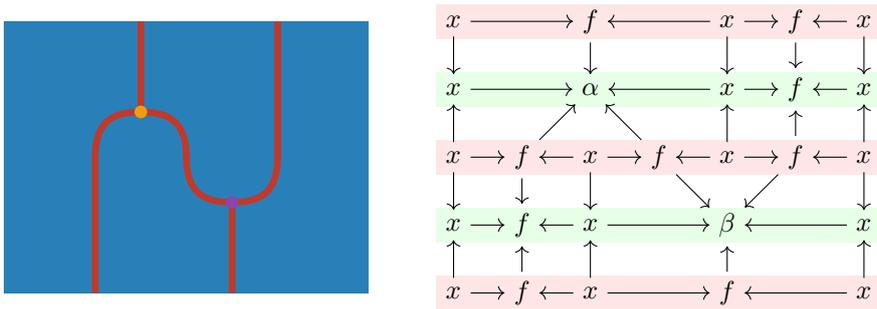

In the implementation, $n$\-cells are represented combinatorially as \emph{$n$\-diagrams}, simple inductive data structures which allow us to represent the mathematical zigzag construction~\cite{reutter2019high}.
We depict this in \Cref{fig:stringdiagzigzag}, which illustrates the encoding of a 2\-diagram (drawn on the left) in terms of an iterated sequence of cospans (drawn on the right).
Here, a zigzag is a sequence of cospans taking value in some category of labels. We distinguish the \emph{singular heights} drawn in green, where vertices might appear, from the \emph{regular heights} drawn in red which are adjacent.
Since zigzags and their morphisms themselves form a category, this construction can be iterated, and in this way higher-dimensional diagrams can be represented.
This inductive nature of the construction is an essential requirement for our proof assistant: it allows us to describe $n$\-cells as algebraic data types, as we will detail in \Cref{sec:datastructures}, and to structure our key algorithms of \Cref{sec:algorithms} as recursive procedures.

\subsection{Related work}

There are many tools for higher categories that use string diagrams as a visualisation method, and we summarise several of them in \Cref{tab:existing-tools}.
Each takes a different categorical perspective, and is focused on a particular formalism.
The most closely related tool is \textsf{Globular}, the direct precursor to \hio, which was limited to 4-categories and lacked full support for coherent inverses.
One other tool, \textsf{rewalt}, also allows manipulations at the level of $n$\-categories for arbitrary $n$.
Both \hio and \textsf{rewalt} implement semistrict $n$\-categories, but they have different notions of semistrictness: \hio has strict associators and unitors, but weak interchanges; whereas \textsf{rewalt} has strict interchangers and associators, with weak identities.
This difference of approach means that the corresponding notions of string diagram are quite different in each tool.
The \hio tool is the first string diagram proof assistant that can handle coherent invertible generators in all dimensions, an aspect it shares with traditional type-theoretic approaches to higher category theory, such as homotopy type theory~\cite{hottbook}, interpreted via proof assistants such as Agda or Coq.

\begin{table}
	\centering
	\begin{tabular}{@{}lllll@{}}
		\toprule
		tool                                                    & generality                         & interactive & invertibility & visualisation \\ \midrule
		\hio                                                    & $n$-categories                    & \cmark      & \cmark        & up to 4D      \\
		\textsf{Cartographer}~\cite{sobocinski2019cartographer} & symmetric monoidal categories      & \cmark      & \xmark        & 2D            \\
		\textsf{DisCoPy}~\cite{de2020discopy}                   & monoidal categories                & \xmark      & \xmark        & 2D            \\
		\textsf{Globular}~\cite{bar2018globular}                & 4-categories                       & \cmark      & partial       & 2D            \\
		\textsf{rewalt}~\cite{hadzihasanovic2022data}           & $n$-categories                    & \xmark      & partial       & 2D            \\
		\textsf{sd-visualiser}~\cite{rice2024sd}                & traced cartesian closed categories & \cmark      & \xmark        & 2D            \\
		\textsf{Quantomatic}~\cite{kissinger2015quantomatic}    & compact closed categories          & \cmark      & \xmark        & 2D            \\
		\textsf{wiggle.py}~\cite{burton2023string}              & monoidal 2-categories              & \xmark      & \xmark        & up to 3D      \\ \bottomrule \\
	\end{tabular}
	\caption{Comparison of existing tools for string diagrams.}
	\label{tab:existing-tools}
\end{table}

This paper is the first detailed description of \hio, with previous works focused on aspects of the theoretical foundations~\cite{reutter2019high, heidemann2022zigzag,tataru2023layout,sarti2023posetal,tataru2024theory,hu2024coherent}.
The theory of associative $n$\-categories, which is the basis of \hio, was first developed by Dorn, Douglas, and Vicary and described in Dorn's PhD thesis~\cite{dorn2018associative}.

\subsection{Acknowledgements}

The authors would like to thank Anastasia Courtney, Yulong Huang, and Jasper Parish for their contributions during their summer internships, Akvil\.e Valentukonyt\.e and Klaudia Urbanska for their contributions during their undergraduate projects, and Manuel Ara\'ujo, Wilf Offord, and Hao Xu for extensive testing of the tool and valuable feedback.
We are also grateful to the students of the \enquote{Categorical Quantum Mechanics} course at Oxford, and the \enquote{Advanced Topics in Category Theory} course at Cambridge for testing and feedback.

\section{Using the tool} \label{sec:usingtool}

The tool consists of two main components: the \emph{signature} and the \emph{workspace}.
The signature stores a list of generators for an $n$\-category, and the workspace stores an $n$\-diagram in the free $n$\-category generated by this signature.
The tools implements a number of \emph{actions} to modify the signature and/or workspace, and every state of the tool is determined by the list of actions that led to that state, starting from the empty signature and workspace.
This makes it easy to implement an undo/redo system, and to reproduce a state by replaying the list of actions which is useful for debugging and testing.

\begin{figure}
	\centering
	\includegraphics[width=\textwidth]{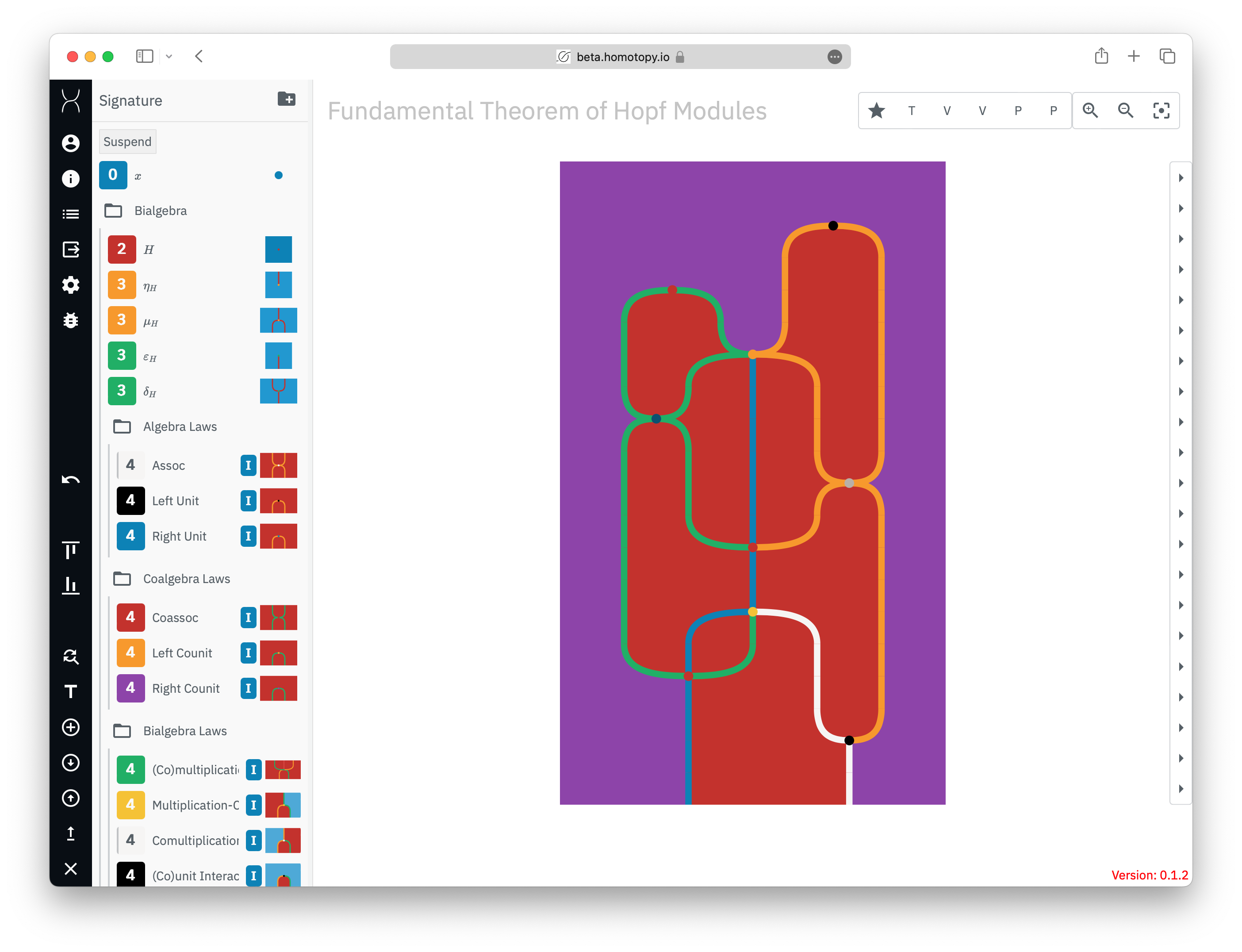}
	\caption{The interface of the proof assistant.}
	\label{fig:userinterface}
\end{figure}

All actions can be triggered by clicking on UI elements such as the buttons on the sidebar, interacting with the workspace diagram, or with keyboard shortcuts.
We denote the keyboard shortcut associated to an action as \keys{A}, which represents pressing the \enquote{A} key.
Unlike a traditional proof assistant, there is no text-based aspect to the user-interface, except for metadata and for providing generator names.
A screenshot of the user interface is shown in \Cref{fig:userinterface}.
In this section, we describe its major components.

\paragraph{Adding generators.}
There are two primary ways to add new generators to the signature:
\begin{itemize}
	\item We can add new 0-dimensional generators by clicking \enquote{Add 0-cell} (\keys{A}) on the sidebar.
	\item We can add new $(n + 1)$\-dimensional generators by constructing a source $n$\-diagram in the workspace and clicking \enquote{Source} (\keys{S}), and then constructing a compatible target $n$\-diagram in the workspace and clicking \enquote{Target} (\keys{T}).
	      The order of these two actions is not important, and the first diagram will be stashed and displayed in the bottom-left corner of the workspace until the second one is constructed.
\end{itemize}

\paragraph{Manipulating the workspace.}
Clicking on any generator in the signature will bring that diagram into the workspace.
We can raise the dimension of the diagram with the \enquote{Identity} (\keys{I}) action.
We can construct composite diagrams by attaching another diagram to a boundary by clicking on the edge of the diagram.
We can also use generators in the signature to perform rewrites by clicking \emph{inside} the workspace diagram.
Finally, we can perform homotopies, such as contractions and expansions which will be described in \Cref{sec:algorithms}, by clicking and dragging.

\paragraph{Theorems.}
If the workspace is displaying an $n$\-diagram $D$ for $n>1$, the \enquote{Theorem} (\keys{H}) action becomes available.
This action creates a new $n$\-dimensional generator $T$ with the same type as $D$, and a new invertible $(n + 1)$\-dimensional generator $P\colon T \to D$. This could be done by hand, and in this sense this action does not strictly add functionality, rather it adds a useful shortcut.

The idea is that $T$ is an algebraic generator which axiomatises the existence of $D$, therefore allowing users to use it as a rewrite, and the generator $P$ witness the fact that $T$ is true (i.e.\ is inhabited), by rewriting it to $D$.
This feature is useful for formalising complex proofs that depend on other lemmas, by saving each lemma as a theorem and then combining them to prove the main theorem.
It can also be used to give definitions---i.e., we can think of $T$ as a new generator which is \emph{defined} to equal $D$, and $P$ can then be used to expand the definition.

\paragraph{View control.}
Visualising $n$\-diagrams for $n > 2$ presents a fundamental challenge, since the geometry can be difficult for us to visualise.
A number of features are therefore provided for manipulating the visualisation: projecting out certain dimensions, navigating to a subdiagram, and changing the rendering dimension (with a choice between 0 and 4 dimensions).

These manipulations are effected via the \emph{view control} component in the top-right corner of the workspace.
This consists of a list of $n + 1$ buttons, where $n$ is the dimension of the workspace diagram, corresponding to the following regular expression:
\[
	\bigstar \; ( \textsf{S} \mid \textsf{T} \mid \textsf{R} i \mid \textsf{S} i )^k \; \textsf{V}^d \; \textsf{P}^{n - k - d} \qquad (d \leq 4)
\]
Here \textsf{S} means source, \textsf{T} means target, $\textsf{R} i$ means the $i$-th regular slice, and $\textsf{S} i$ means the $i$-th singular slice.
This means that we are viewing a $k$\-dimensional subdiagram of the workspace diagram, which is an $(n - k)$\-diagram obtained by recursively going into slices as specified by the first $k$ symbols.
This subdiagram is then projected to $d$ dimensions; that is, we are \emph{viewing} $d$ dimensions, and \emph{projecting} the remaining $n - k - d$ dimensions.
Clicking the star will reset the slice to the original $n$\-diagram (i.e.\ reset $k$ to $0$), and clicking on any of the $k$ slice buttons will reset the slice to some prefix.
Clicking on any \textsf{V} button will decrement $d$ by 1, to a minimum of 0 (displayed as a single point) and clicking on any \textsf{P} button will increment $d$ by 1, to a maximum of $4$ (rendered as a movie of 3D geometries).

Users can descend into a slice by clicking the chevrons appearing on the right-hand side of the diagram, or descend into the source slice by pressing \keys{\arrowkeyright}.
Similarly, while inside a slice, users can ascend to the parent slice by pressing \keys{\arrowkeyleft}, and navigate to adjacent $k$\-th slice components by pressing \keys{\arrowkeyup}/\keys{\arrowkeydown}.

To illustrate how the projection works, consider \Cref{fig:associator} which shows the 2D and 3D projections of a 3-dimensional diagram side-by-side.
Note that the 2D projection can be understood as looking at the 3D projection from below, and projecting onto a 2D plane.

\paragraph{Generator options.}
When hovering over a generator, a \faCog\ icon appears on the left which reveals a menu of options for that generator.
This allows the user to rename the generator (with support for \LaTeX) or change its colour or shape.
Most importantly, it allows the user to mark a generator as invertible.

\paragraph{Image export.}
The sidebar has an \enquote{Image Export} button which reveals a panel for exporting the workspace diagram to different formats, such as SVG, TikZ, Manim, and STL.

\section{Example: Eckmann-Hilton} \label{sec:example}

Here we will illustrate how the proof assistant may be used in practice to formalise results in higher category theory.
Our running example will be the Eckmann-Hilton argument, the key result in the correspondence between braided monoidal categories and 3-categories which are doubly-degenerate, meaning they have a unique 0-cell and no non-identity 1-cells.
This will be essential in \Cref{sec:casestudy} for our formalisation of Hopf algebras in braided monoidal categories.
This section is accompanied by a video tutorial, which can be watched on YouTube at \url{https://www.youtube.com/watch?v=ceu20ahau-I}.
The resulting workspace can be loaded into the tool at \url{https://beta.homotopy.io/p/2402.00002}.

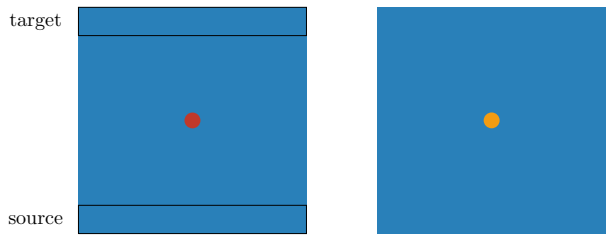
\begin{figure}
	\centering
	\scalebox{0.75}{\begin{tabular}{c  c}
\begin{tabular}{@{}c@{}}
\begin{tikzpicture}
\definecolor{generator-0-0-0-pos}{RGB}{41, 128, 185}
\definecolor{generator-1-2-0-pos}{RGB}{192, 57, 43}
\begin{scope}
\fill[generator-0-0-0-pos] (0,0) -- (4,0) -- (4,4) -- (0,4) -- (0,0);
\draw (0,3.5) -- (4,3.5) -- (4,4) -- (0,4) -- (0,3.5);
\draw (0,0) -- (4,0) -- (4,0.5) -- (0,0.5) -- (0,0);
\end{scope}
\fill[generator-1-2-0-pos] (2,2) circle (0.14);
\node at (-0.75, 0.25) {source};
\node at (-0.75, 3.75) {target};
\end{tikzpicture}
\end{tabular}
\qquad & \qquad
\begin{tabular}{@{}c@{}}
\begin{tikzpicture}
\definecolor{generator-0-0-0-pos}{RGB}{41, 128, 185}
\definecolor{generator-2-2-0-pos}{RGB}{243, 156, 18}
\begin{scope}
\fill[generator-0-0-0-pos] (0,0) -- (4,0) -- (4,4) -- (0,4) -- (0,0);
\end{scope}
\fill[generator-2-2-0-pos] (2,2) circle (0.14);
\end{tikzpicture}
\end{tabular}
\end{tabular}}
	\caption{The scalars $\alpha$ and $\beta$ in our signature for the Eckmann-Hilton argument.}%
	\label{fig:ehsignature}
\end{figure}

To formalise the Eckmann-Hilton argument, we load the tool and begin constructing our signature, which is given by a unique 0-cell $x$ and two 2-cells $\alpha$, $\beta$ which have source and target the identity on $x$.
Since $x$ is a 0-cell, we may immediately add this to the signature by pressing the \enquote{Add 0-Cell} button, and renaming the cell to \enquote{\texttt{\$x\$}}---the use of \enquote{\texttt{\$}}s is optional, but indicates to the tool that the generator's name should be interpreted as \LaTeX.
To add the two scalars $\alpha$ and $\beta$ to the signature, we need to provide the tool with their sources and targets, which in this case is $1_x$.
Selecting the generator $x$ we have added to the signature, we make it the current working diagram.
We then click on the \enquote{Identity} button to get $1_x$ from $x$, and take it as the source of a new 2-cell $\alpha$ by clicking on \enquote{Source}.
Repeating the construction of $1_x$, we can provide the tool with the target of $\alpha$.
Since the diagrams have compatible boundaries, the globularity condition is satisfied and the tool allows to to click on \enquote{Target} to create a 2-cell, which we rename to \enquote{\texttt{\$\textbackslash alpha\$}}.
By repeating this procedure for $\beta$, we get the signature depicted in \Cref{fig:ehsignature}.

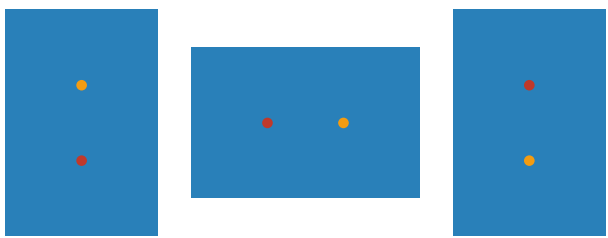
\begin{figure}
	\centering
	\scalebox{0.5}{\definecolor{generator-2-2-0-pos}{RGB}{243, 156, 18}
\definecolor{generator-0-0-0-pos}{RGB}{41, 128, 185}
\definecolor{generator-1-2-0-pos}{RGB}{192, 57, 43}

\begin{tabular}{c c c}
\begin{tabular}{@{}c@{}}
\begin{tikzpicture}
\begin{scope}
\fill[generator-0-0-0-pos] (0,0) -- (4,0) -- (4,6) -- (0,6) -- (0,0);
\end{scope}
\fill[generator-1-2-0-pos] (2,2) circle (0.14);
\fill[generator-2-2-0-pos] (2,4) circle (0.14);
\end{tikzpicture}
\end{tabular}
\quad & \quad
\begin{tabular}{@{}c@{}}
\begin{tikzpicture}
\begin{scope}
\fill[generator-0-0-0-pos] (0,0) -- (6,0) -- (6,4) -- (0,4) -- (0,0);
\end{scope}
\fill[generator-1-2-0-pos] (2,2) circle (0.14);
\fill[generator-2-2-0-pos] (4,2) circle (0.14);
\end{tikzpicture}
\end{tabular}
\quad & \quad
\begin{tabular}{@{}c@{}}
\begin{tikzpicture}
\begin{scope}
\fill[generator-0-0-0-pos] (0,0) -- (4,0) -- (4,6) -- (0,6) -- (0,0);
\end{scope}
\fill[generator-2-2-0-pos] (2,2) circle (0.14);
\fill[generator-1-2-0-pos] (2,4) circle (0.14);
\end{tikzpicture}
\end{tabular}
\end{tabular}}
	\caption{The Eckmann-Hilton argument.}%
	\label{fig:ehproof1}
\end{figure}

With our signature in place, we can begin proving facts about the compositional behaviour of our scalars.
The essence of the Eckmann-Hilton argument amounts to the commutativity and coincidence of vertical and horizontal composites of scalars, as depicted in \Cref{fig:ehproof1}.

Let us detail how this proof can be built.
We begin by constructing the right-most diagram in \Cref{fig:ehproof1} by selecting $\alpha$ from the signature, clicking on the region marked as \emph{target} in \Cref{fig:ehsignature} and selecting $\beta$ to attach.
This builds the vertical composite of $\alpha$ and $\beta$ and gives us the starting point or source for our proof.
Since we wish to collect the steps of the proof into a 3-cell, we take the identity on the current diagram.
This does not appear to change the current diagram displayed but adds an extra dimension in the view control.

We then drag $\beta$ towards $\alpha$ along the right to get the middle diagram, triggering a \emph{contraction} procedure, which we will detail in \Cref{subsec:contraction}.
Next, we drag $\beta$ further downwards, triggering the dual procedure of \emph{expansion}, detailed in \Cref{subsec:expansion}.
Note that these procedures only succeed if they are sound, i.e.\ the resulting diagram has a valid type as checked by the procedure of \Cref{subsec:typechecking}.

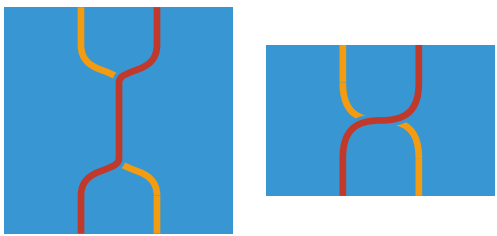
\begin{figure}
	\centering
	\scalebox{0.5}{\definecolor{generator-0-0-1-pos}{RGB}{56, 150, 211}
\definecolor{generator-1-2-0-pos}{RGB}{192, 57, 43}
\definecolor{generator-2-2-0-pos}{RGB}{243, 156, 18}

\newcommand{\wire}[2]{
  \ifdefined\recolor\draw[color=\recolor, line width=10pt]\else\draw[color=#1, line width=5pt]\fi #2
}
\newcommand{\clipped}[3]{
\begin{scope}
  \newcommand{\recolor}{#1}
  \clip#3;
  #2
\end{scope}
}

\begin{tabular}{c c}
\begin{tabular}{@{}c@{}}
\begin{tikzpicture}
\begin{scope}[transparency group]
\fill[generator-0-0-1-pos] (0,0) -- (6,0) -- (6,6) -- (0,6) -- (0,0);
\newcommand{\layer}[1]{
  \clipped{generator-0-0-1-pos}{#1}{(0,0) -- (6,0) -- (6,6) -- (0,6) -- (0,0)}
  #1
}

\wire{generator-2-2-0-pos}{(4,1) .. controls (4,1.8) and (3,1.6) .. (3,2)(3,4) .. controls (3,4.4) and (2,4.2) .. (2,5)};
\layer{
\wire{generator-1-2-0-pos}{(2,0) -- (2,1) .. controls (2,1.8) and (3,1.6) .. (3,2) -- (3,4) .. controls (3,4.4) and (4,4.2) .. (4,5) -- (4,6)};
\wire{generator-2-2-0-pos}{(4,0) -- (4,1)(2,5) -- (2,6)};
}
\end{scope}
\end{tikzpicture}
\end{tabular}
\quad & \quad
\begin{tabular}{@{}c@{}}
\begin{tikzpicture}
\begin{scope}[transparency group]
\fill[generator-0-0-1-pos] (0,0) -- (6,0) -- (6,4) -- (0,4) -- (0,0);
\newcommand{\layer}[1]{
  \clipped{generator-0-0-1-pos}{#1}{(0,0) -- (6,0) -- (6,4) -- (0,4) -- (0,0)}
  #1
}

\wire{generator-2-2-0-pos}{(4,1) .. controls (4,1.8) and (3.6,2) .. (3,2) .. controls (2.4,2) and (2,2.2) .. (2,3)};
\layer{
\wire{generator-1-2-0-pos}{(2,0) -- (2,1) .. controls (2,1.8) and (2.4,2) .. (3,2) .. controls (3.6,2) and (4,2.2) .. (4,3) -- (4,4)};
\wire{generator-2-2-0-pos}{(4,0) -- (4,1)(2,3) -- (2,4)};
}
\end{scope}
\end{tikzpicture}
\end{tabular}
\end{tabular}}
	\caption{The Eckmann-Hilton proof (left) and its contraction (right).}%
	\label{fig:ehproof2}
\end{figure}

We now inspect the proof from the top-dimension by clicking on the $\star$ button on the right.
From this view, our proof amounts to the construction of a braid as depicted in the first image of \Cref{fig:ehproof2}.
We can see this more clearly if we contract the middle part of the diagram by vertically dragging the bottom half-braid towards the top, obtaining the second image of \Cref{fig:ehproof2}.
We may also perform the proof entirely from this view, by horizontally dragging the legs of the braid past each other.
A 3D visualisation of this proof can now be observed by pressing the \textsf{P} button in the slice controls.

\section{Core data structures} \label{sec:datastructures}

In this section we outline the fundamental algebraic data types which are used the encode $n$\-diagrams in the implementation.
These are diagrams, rewrites, cones and cospans; their definitions are sketched in \Cref{fig:core-data-types}.

\begin{figure}[h]
\centering
\begin{lstlisting}[language={OCaml}]
type frame = int
type generator = { dimension: int; id: int }

type rewrite =
	| Rewrite0Identity
	| Rewrite0 of { source: generator; target: generator; label: frame }
	| RewriteN of { cones: cone list }

and cone = {
	index: int;
	source: cospan list;
	target: cospan;
	slices: rewrite list;
}

and cospan = { forward: rewrite; backward: rewrite }

type diagram =
	| Diagram0 of generator
	| DiagramN of { source: diagram; cospans: cospan list }
\end{lstlisting}
\caption{The core data structures of \hio.}
\label{fig:core-data-types}
\end{figure}


\subsection{Diagrams}

The core data structure of \hio is a recursive encoding for diagrams that is derived from the zigzag construction.
A 0-diagram has trivial shape (a single point), and type is essentially the labelling which is an assignment of name to colour, which we call a \emph{generator}.
An $(n + 1)$\-diagram is determined by an alternating sequence of singular and regular slices, which are $n$\-diagrams, together with information on how the slices fit together:
\[
	\begin{tikzcd}
		{r_0} \ar[r, "f_0"] &
		{s_1} &
		{r_1} \ar[l, "b_0"'] \ar[r, "f_1"] &
		{s_1} &
		{r_1} \ar[l, "b_1"'] \ar[r, "f_2"] &
		{s_2} &
		{r_2}. \ar[l, "b_2"']
	\end{tikzcd}
\]
Each singular slice $s_i$ is equipped with rewrites $f_i\colon r_i \to s_i$ and $b_i\colon r_{i + 1} \to s_i$ that encode the difference between $s_i$ and its neighbouring regular slices $r_i$ and $r_{i + 1}$.
We call the pair $(f_i, b_i)$ a cospan, and describe the height of a diagram as the number of cospans (equally, singular slices) it contains.
There is always one more regular slice than singular slices.
Instead of storing all the slices of a diagram, we only keep around the first regular slice $r_0$ and reconstruct the other slices when needed by applying the rewrites backwards and forwards.

\subsection{Rewrites}

\begin{figure}
	\centering
	\scalebox{0.75}{\begin{tikzpicture}
	\definecolor{generator-0-0-0-pos}{RGB}{41, 128, 185}
	\definecolor{generator-1-1-0-pos}{RGB}{192, 57, 43}
	\definecolor{generator-2-2-0-pos}{RGB}{243, 156, 18}
	\begin{scope}
		\fill[generator-0-0-0-pos] (0,0) -- (8,0) -- (8,4) -- (0,4) -- (0,0);
		\draw[color=generator-1-1-0-pos, line width=5pt](2,0) -- (2,4)(4,0) -- (4,1) .. controls (4,1.8) and (4.4,2) .. (5,2) -- (5,4)(6,0) -- (6,1) .. controls (6,1.8) and (5.6,2) .. (5,2);
	\end{scope}
	\fill[generator-2-2-0-pos] (5,2) circle (0.14);
	\draw[fill=yellow, opacity=0.2] (1.5,0) -- (2,2) -- (2.5,0) -- cycle node [opacity=1, midway, below] {$A$};
	\draw[fill=yellow, opacity=0.2] (1.5,4) -- (2,2) -- (2.5,4) -- cycle node [opacity=1, midway, above] {$B$};
	\draw[fill=yellow, opacity=0.2] (3.5,0) -- (5,2) -- (6.5,0) -- cycle node [opacity=1, midway, below] {$C$};
	\draw[fill=yellow, opacity=0.2] (4.5,4) -- (5,2) -- (5.5,4) -- cycle node [opacity=1, midway, above] {$D$};
\end{tikzpicture}}
	\caption{A zigzag with four cones. $A$ and $B$ are \emph{identity} cones that do not change the diagram, omitted in the sparse representation.}%
	\label{fig:cones}
\end{figure}
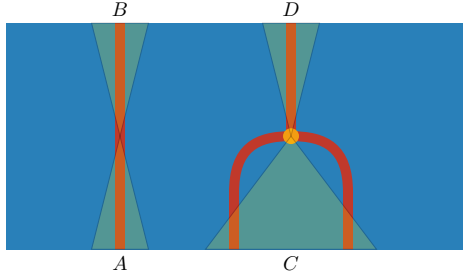

Rewrites between diagrams are also encoded as a recursive data structure.
A rewrite of 0-diagrams (0-rewrite) is either the identity rewrite or a rewrite between the underlying generators; in the latter case, it is also equipped with a \emph{frame} which represents the directionality of the rewrite in some space associated to the ambient diagram.
A $(n+1)$\-rewrite $x \to y$ modifies the sequence of cospans in $x$ by removing subsequences and replacing them with individual cospans.
Each such modification is called a \emph{cone} (see \Cref{fig:cones}).
The following diagram, which encodes the bottom half of \Cref{fig:cones}, is an example of a rewrite with a single cone that replaces the cospans $(f_1, b_1)$ and $(f_2, b_2)$ with the cospan $(f^\prime, b^\prime)$:
\[
	\begin{tikzcd}
		{r_0} \ar[r, "f_0"] &
		{s_1} &
		{r_1} \ar[l, "b_0"'] \ar[rr, "f^\prime"] &
		{} &
		{s_1^\prime} &
		{} &
		{r_2} \ar[ll, "b^\prime"'] \\
		{r_0} \ar[r, "f_0"'] \ar[u, dashed] &
		{s_1} \ar[u, dashed] &
		{r_1} \ar[l, "b_0"] \ar[r, "f_1"'] \ar[u, dashed] &
		{s_1} \ar[ur, "\ell_1"] &
		{r_1} \ar[l, "b_1"] \ar[r, "f_2"'] &
		{s_2} \ar[ul, "\ell_2"'] &
		{r_2} \ar[l, "b_2"] \ar[u, dashed]
	\end{tikzcd}
\]
A cone in a rewrite of $(n + 1)$\-diagrams also contains rewrites between the $n$\-dimensional singular slices.
In the example above these are the rewrites $\ell_1\colon s_1 \to s_1^\prime$ and $\ell_2\colon s_2 \to s_1^\prime$.
A rewrite does not store information for the parts of the diagrams that do not change.
Since in practice most adjacent slices only differ in small parts, this sparse encoding leads to significant space efficiency.

The core data structures admit a series of auxiliary algorithms:
\begin{enumerate}
	\item Given an $n$-diagram $x$ we can apply a rewrite $x \xrightarrow{R} y$ forwards to obtain $y$.
	      Similarly, given an $n$-diagram $y$ and a rewrite $x \xrightarrow{R} y$ we can apply $R$ backwards to $y$.
	\item We can compute the slices of an $(n + 1)$\-diagram at any singular or regular height. In particular, we can obtain the source and target of a diagram.
	\item When $x \xrightarrow{R} y$ and $y \xrightarrow{R^\prime} z$ are rewrites, we can compute the composite rewrite $x \xrightarrow{R^\prime \circ R} z$
	      while preserving sparsity.
	\item For two $n$\-diagrams $s$ and $t$ whose sources and targets agree, we can create an $(n + 1)$-dimensional diagram which represents an $(n + 1)$-morphism between $s$ and $t$.
	\item \label{alg:projection} Given an $n$\-diagram $x$ and $k \leq n$ we can compute the generators that would be visible in the projection of $x$ to $k$ dimensions.
	\item \label{alg:explosion} For an $n$\-diagram $x$ and $k \leq n$ we can produce a graph that is a dense encoding of the projection of $x$ to $k$ dimensions. The simplicial complex obtained as the flag complex from this graph has the geometry of the rendered diagram. The layout algorithm then assigns coordinates to the vertices (see~\Cref{subsec:layout}).
	\item Given a pair of diagrams $x$, $y$ we can search for copies of $y$ that are embedded into $x$ and intersect a line through the projection. This allows us to find opportunities to rewrite a diagram, which itself can be realised as a higher cell.
\end{enumerate}

\section{Key algorithms}\label{sec:algorithms}

Many of our key algorithms described in this section first compute the shape of the result, usually by recursion over diagram dimension, and then complete the type information.
These data are encoded by directed graphs, for which we can utilise graph algorithms to implement our operations.

\subsection{Collapse}\label{subsec:collapse}

A string diagram has geometric properties (e.g.\ the length of a wire) which are not intended to be meaningful, as its meaning is captured entirely by topological properties (e.g.\ connectivity of a wire).
Topology yields a natural compatibility with composition in the string diagram calculus: if two wires are placed in sequence, then the resulting string diagram is merely a longer wire, and therefore it looks topologically the same and represents the same mathematical content.
However, both the tool and the theory are based on \emph{combinatorial} encodings of string diagrams, and moreover there is a distinct encoding for the diagram of two wires in sequence versus one wire.

We describe the \emph{collapse} of a diagram as a combinatorial representation of these topological invariants, essentially as a directed graph, which gives a normal form for the fully-exploded graph obtained from an $n$-diagram in a frame-preserving way.
In more detail, given such a graph, every node has a neighbourhood which determines a set of outgoing frames and incoming frames (framing data); now, consider the largest equivalence relation on nodes determined by $x \sim y$ when $x$ is adjacent to $y$ by an identity 0-rewrite, and $x$ and $y$ admit equal framing data; collapse is the quotient graph induced by $\sim$.
Explicitly, we compute this by treating the graph as a simplicial complex and checking each 1-simplex (edge) which is an identity 0-rewrite to see if it is collapsible, whereby identifying both 0-simplex faces (endpoints) respects $\sim$, by checking all 2-simplices (triangles) for which that 1-simplex is a face.

Collapse is used to compare when two diagrams differ as encodings but have the same topological data: many different diagrams may have the same collapse, for instance degenerating a diagram along any part\footnote{This corresponds to composition with a weak unit in the $n$-categorical model; combinatorially, this represents the insertion of redundant data.} does not change its collapse, but two diagrams with distinct collapses are necessarily distinct.
These degeneracies arise naturally in the course of the other operations described in this section, but can be often eliminated without rendering the diagram malformed.
Collapse is a crucial component of typechecking, analogously to how the standard technique of deciding equality in a term calculus often amounts to computing normal forms for each term.
A detailed account is given in~\cite{hu2024coherent}.

\subsection{Contraction} \label{subsec:contraction}

\emph{Contraction} is the process of shortening a diagram along a homotopy, producing a contraction rewrite $D \xrightarrow{c} C$, which may or may not exist depending on $D$.
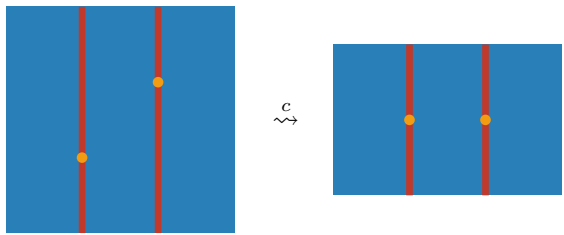
\begin{figure}
	\[
		\vcenter{\hbox{\scalebox{0.5}{\begin{tikzpicture}
\definecolor{generator-0-0-0-pos}{RGB}{41, 128, 185}
\definecolor{generator-1-1-0-pos}{RGB}{192, 57, 43}
\definecolor{generator-2-2-0-pos}{RGB}{243, 156, 18}
\begin{scope}
\fill[generator-0-0-0-pos] (0,0) -- (6,0) -- (6,6) -- (0,6) -- (0,0);
\draw[color=generator-1-1-0-pos, line width=5pt](2,0) -- (2,6)(4,0) -- (4,6);
\end{scope}
\fill[generator-2-2-0-pos] (2,2) circle (0.14);
\fill[generator-2-2-0-pos] (4,4) circle (0.14);
\end{tikzpicture}}}}
		\quad\overset{c}{\rightsquigarrow}\quad
		\vcenter{\hbox{\scalebox{0.5}{\begin{tikzpicture}
\definecolor{generator-0-0-0-pos}{RGB}{41, 128, 185}
\definecolor{generator-1-1-0-pos}{RGB}{192, 57, 43}
\definecolor{generator-2-2-0-pos}{RGB}{243, 156, 18}
\begin{scope}
\fill[generator-0-0-0-pos] (0,0) -- (6,0) -- (6,4) -- (0,4) -- (0,0);
\draw[color=generator-1-1-0-pos, line width=5pt](2,0) -- (2,4)(4,0) -- (4,4);
\end{scope}
\fill[generator-2-2-0-pos] (2,2) circle (0.14);
\fill[generator-2-2-0-pos] (4,2) circle (0.14);
\end{tikzpicture}}}}
	\]
	\caption{A contraction of two beads on two singular heights.}%
	\label{fig:bead-contraction}
\end{figure}
An example of this is given by \Cref{fig:bead-contraction}.
The result of the contraction $C$ always has a singular height of one.
Informally, as a homotopy, the contraction rewrite $c$ is \emph{the} unique canonical way to shorten the diagram without making any arbitrary choices, and this is mathematically captured by its description as the computation of a categorical colimit.

Contraction works recursively on diagram dimension, with the 0-dimensional base case obtained via collapse, as in \Cref{subsec:collapse}, and then ensuring that the maximal elements of the poset induced by the reachability relation on the resulting directed graph are compatible~\cite{hu2024coherent}.
The higher-dimensional recursive case has been described theoretically by Reutter and Vicary~\cite{reutter2019high}; it works by first determining a \enquote{$\Delta$-colimit}, which determines the shape that the result necessarily possesses, and then if that exists it then attempts to find a compatible labelling of each stratum to complete the typing information by a divide-and-conquer strategy.

In the base case, compatibility is the uniqueness of the labelling of the maximal element of the poset, combined with the condition that the framing data on each maximal element is identical. In the recursive case, for an $(n+1)$-rewrite $D \xrightarrow{c} C$, the algorithm is as follows:
\begin{enumerate}
	\item $D$ is associated to a directed graph $G$ whose nodes represent regular and singular slices of $D$, weighted by $n$-diagrams, and whose edges form cospans between regular slices and are weighted by $n$-rewrites;
	\item \label{alg:contraction-recursive} this directed graph is then exploded, obtaining a larger directed graph $E$ whose nodes are weighted by $(n-1)$-diagrams, and whose edges are weighted by $(n-1)$-rewrites, by replacing each node of $G$ with a directed graph, as in the previous step, and each edge by a collection of connecting $(n-1)$-rewrites;
	\item $E$ induces a subgraph $\Delta_E$ which represents only the shape data of $D$ consisting of only the singular-singular nodes; on $\Delta_E$, we obtain its condensation graph with respect to strongly-connected components\footnote{This graph has a node for each strongly-connected component, and each edge represents reachability for strongly-connected components.}; if the edge relation of the condensation graph does not correspond to a linear order on nodes, then fail, otherwise, this determines:
	      \begin{enumerate}
		      \item subproblems determined by the subgraph of $E$ induced by reachability for the nodes of a particular strongly-connected component of $\Delta_E$;
		      \item a linear ordering of subproblems, which will determine how their solutions should be combined.
	      \end{enumerate}
	\item each subproblem represents an independent fragment of $D$ which spans its entire height, and is solved recursively (step \labelcref{alg:contraction-recursive}); its solution is a singular-singular slice of $C$, which necessitates the linear ordering of subproblems; the rest of $C$ and $c$ are constructed from the remaining data;
	\item the result is then typechecked, and if it fails, then the algorithm fails.
\end{enumerate}

There are also auxiliary algorithms which propagate a contraction within some slice of a larger diagram.
We refer the interested reader to~\cite{reutter2019high}.

\subsection{Expansion} \label{subsec:expansion}

\emph{Expansion} is the dual of contraction, first described by Reutter and Vicary~\cite{reutter2019high}, and later refined by Tataru and Vicary~\cite{tataru2024theory}.
It takes a diagram $D$ and produces an expanded diagram $E$, such that $E$ in fact contracts to give $D$ via some rewrite $E \xrightarrow{c} D$. In this sense, expansion is a partial converse to contraction.

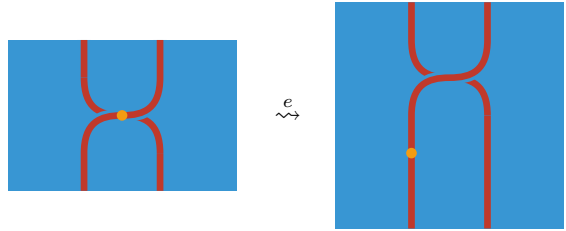
\begin{figure}
    \centering
    \begin{equation*}
    	\vcenter{\hbox{\scalebox{0.5}{\begin{tikzpicture}
\definecolor{generator-0-0-1-pos}{RGB}{56, 150, 211}
\definecolor{generator-1-2-0-pos}{RGB}{192, 57, 43}
\definecolor{generator-2-3-0-pos}{RGB}{243, 156, 18}

\newcommand{\wire}[2]{
  \ifdefined\recolor\draw[color=\recolor, line width=10pt]\else\draw[color=#1, line width=5pt]\fi #2
}
\newcommand{\clipped}[3]{
\begin{scope}
  \newcommand{\recolor}{#1}
  \clip#3;
  #2
\end{scope}
}

\begin{scope}[transparency group]
\fill[generator-0-0-1-pos] (0,0) -- (6,0) -- (6,4) -- (0,4) -- (0,0);
\newcommand{\layer}[1]{
  \clipped{generator-0-0-1-pos}{#1}{(0,0) -- (6,0) -- (6,4) -- (0,4) -- (0,0)}
  #1
}

\wire{generator-1-2-0-pos}{(4,1) .. controls (4,1.8) and (3.6,2) .. (3,2) .. controls (2.4,2) and (2,2.2) .. (2,3)};
\layer{
\wire{generator-1-2-0-pos}{(2,0) -- (2,1) .. controls (2,1.8) and (2.4,2) .. (3,2) .. controls (3.6,2) and (4,2.2) .. (4,3) -- (4,4)(4,0) -- (4,1)(2,3) -- (2,4)};
}
\end{scope}
\fill[generator-2-3-0-pos] (3,2) circle (0.14);
\end{tikzpicture}}}}
        \quad\overset{e}{\rightsquigarrow}\quad
    	\vcenter{\hbox{\scalebox{0.5}{\begin{tikzpicture}
\definecolor{generator-0-0-1-pos}{RGB}{56, 150, 211}
\definecolor{generator-1-2-0-pos}{RGB}{192, 57, 43}
\definecolor{generator-2-3-0-pos}{RGB}{243, 156, 18}

\newcommand{\wire}[2]{
  \ifdefined\recolor\draw[color=\recolor, line width=10pt]\else\draw[color=#1, line width=5pt]\fi #2
}
\newcommand{\clipped}[3]{
\begin{scope}
  \newcommand{\recolor}{#1}
  \clip#3;
  #2
\end{scope}
}

\begin{scope}[transparency group]
\fill[generator-0-0-1-pos] (0,0) -- (6,0) -- (6,6) -- (0,6) -- (0,0);
\newcommand{\layer}[1]{
  \clipped{generator-0-0-1-pos}{#1}{(0,0) -- (6,0) -- (6,6) -- (0,6) -- (0,0)}
  #1
}

\wire{generator-1-2-0-pos}{(4,3) .. controls (4,3.8) and (3.6,4) .. (3,4) .. controls (2.4,4) and (2,4.2) .. (2,5)};
\layer{
\wire{generator-1-2-0-pos}{(2,0) -- (2,3) .. controls (2,3.8) and (2.4,4) .. (3,4) .. controls (3.6,4) and (4,4.2) .. (4,5) -- (4,6)(4,0) -- (4,3)(2,5) -- (2,6)};
}
\end{scope}
\fill[generator-2-3-0-pos] (2,2) circle (0.14);
\end{tikzpicture}}}}
    \end{equation*}
    \caption{An anticontraction move.}
    \label{fig:anticontraction}
\end{figure}

It is defined inductively, similarly to contraction.
In the base case, it performs an interchanger move that separates two singular levels (the reverse of \Cref{fig:bead-contraction}). In the recursive case, expansion attempts to propagate an expansion of a sub-diagram to the diagram itself, yielding a diagram which will contract to the original.
Since contraction is computed by a colimit process, this requires an algorithm that can reverse the ordinary colimit process, a procedure that we call \emph{anticontraction}~\cite{tataru2024theory}.
We illustrate this with the recursive expansion example in \Cref{fig:anticontraction}, where a vertex is moved out of a singular height. Note that the expanded diagram on the right indeed contracts to give the original diagram on the left.

\subsection{Typechecking} \label{subsec:typechecking}

Typechecking is the process of checking the validity of a diagram with respect to a given signature.
The underlying theory gives a clear perspective on this process~\cite{dorn2018associative,heidemann2022zigzag}: break the diagram into atomic \enquote{pieces}, and ensure that each piece collapses to the canonical neighbourhood of the respective generator.
This tells us that the neighbourhood of each point in the diagram is fully described by the signature.

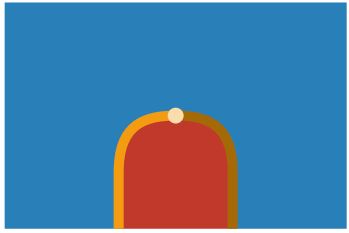
\begin{figure}
	\centering
	\scalebox{0.75}{\begin{tikzpicture}
\definecolor{generator-1-0-0-pos}{RGB}{192, 57, 43}
\definecolor{generator-2-1-0-neg}{RGB}{166, 105, 8}
\definecolor{generator-2-1-0-pos}{RGB}{243, 156, 18}
\definecolor{generator-0-0-0-pos}{RGB}{41, 128, 185}
\definecolor{generator-2-1-1-zer}{RGB}{251, 221, 173}
\begin{scope}
\fill[generator-0-0-0-pos] (0,0) -- (2,0) -- (2,1) .. controls (2,1.8) and (2.4,2) .. (3,2) .. controls (3.6,2) and (4,1.8) .. (4,1) -- (4,0) -- (6,0) -- (6,4) -- (0,4) -- (0,0);
\fill[generator-1-0-0-pos] (2,0) -- (4,0) -- (4,1) .. controls (4,1.8) and (3.6,2) .. (3,2) .. controls (2.4,2) and (2,1.8) .. (2,1) -- (2,0);
\draw[color=generator-2-1-0-neg, line width=5pt](4,0) -- (4,1) .. controls (4,1.8) and (3.6,2) .. (3,2);
\draw[color=generator-2-1-0-pos, line width=5pt](2,0) -- (2,1) .. controls (2,1.8) and (2.4,2) .. (3,2);
\end{scope}
\fill[generator-2-1-1-zer] (3,2) circle (0.14);
\end{tikzpicture}}
	\caption{The cancellation 2-cell $f^{-1} \circ f \to 1_x$ generated by an invertible 1-cell $f\colon x \to y$.}
	\label{fig:counit}
\end{figure}

Collapse plays a crucial role here, because it ensures that higher-dimensional coherences of invertible generators also typecheck.
For example, if $f\colon x \to y$ is an invertible 1-cell, we can generate a 2-cell $f^{-1} \circ f \to 1_x$ known as the \emph{counit} (see \Cref{fig:counit}); this is well-typed because it collapses to the canonical neighbourhood of $f$.

\subsection{Layout} \label{subsec:layout}

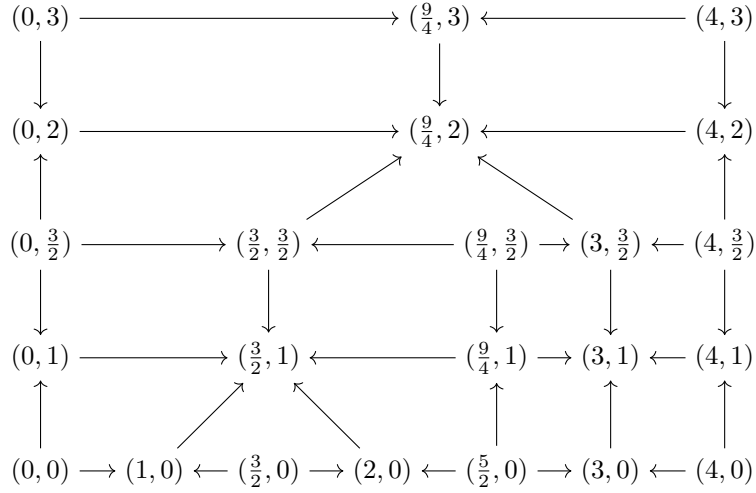
\begin{figure}
	\centering
	\begin{tikzpicture}[scale=1.5]
\node (R0R0) at (0, 0) {$(0, 0)$};
\node (R0S0) at (1, 0) {$(1, 0)$};
\node (R0R1) at (2, 0) {$(\frac{3}{2}, 0)$};
\node (R0S1) at (3, 0) {$(2, 0)$};
\node (R0R2) at (4, 0) {$(\frac{5}{2}, 0)$};
\node (R0S2) at (5, 0) {$(3, 0)$};
\node (R0R3) at (6, 0) {$(4, 0)$};

\node (S0R0) at (0, 1) {$(0, 1)$};
\node (S0S0) at (2, 1) {$(\frac{3}{2}, 1)$};
\node (S0R1) at (4, 1) {$(\frac{9}{4}, 1)$};
\node (S0S1) at (5, 1) {$(3, 1)$};
\node (S0R2) at (6, 1) {$(4, 1)$};

\node (R1R0) at (0, 2) {$(0, \frac{3}{2})$};
\node (R1S0) at (2, 2) {$(\frac{3}{2}, \frac{3}{2})$};
\node (R1R1) at (4, 2) {$(\frac{9}{4}, \frac{3}{2})$};
\node (R1S1) at (5, 2) {$(3, \frac{3}{2})$};
\node (R1R2) at (6, 2) {$(4, \frac{3}{2})$};

\node (S1R0) at (0, 3) {$(0, 2)$};
\node (S1S0) at (3.5, 3) {$(\frac{9}{4}, 2)$};
\node (S1R1) at (6, 3) {$(4, 2)$};

\node (R2R0) at (0, 4) {$(0, 3)$};
\node (R2S0) at (3.5, 4) {$(\frac{9}{4}, 3)$};
\node (R2R1) at (6, 4) {$(4, 3)$};

\draw[->] (R0R0) to (R0S0);
\draw[->] (R0R1) to (R0S0);
\draw[->] (R0R1) to (R0S1);
\draw[->] (R0R2) to (R0S1);
\draw[->] (R0R2) to (R0S2);
\draw[->] (R0R3) to (R0S2);

\draw[->] (S0R0) to (S0S0);
\draw[->] (S0R1) to (S0S0);
\draw[->] (S0R1) to (S0S1);
\draw[->] (S0R2) to (S0S1);

\draw[->] (R1R0) to (R1S0);
\draw[->] (R1R1) to (R1S0);
\draw[->] (R1R1) to (R1S1);
\draw[->] (R1R2) to (R1S1);

\draw[->] (S1R0) to (S1S0);
\draw[->] (S1R1) to (S1S0);

\draw[->] (R2R0) to (R2S0);
\draw[->] (R2R1) to (R2S0);

\draw[->] (R0R0) to (S0R0);
\draw[->] (R0R2) to (S0R1);
\draw[->] (R0R3) to (S0R2);

\draw[->] (R0S0) to (S0S0);
\draw[->] (R0S1) to (S0S0);
\draw[->] (R0S2) to (S0S1);

\draw[->] (R1R0) to (S0R0);
\draw[->] (R1R1) to (S0R1);
\draw[->] (R1R2) to (S0R2);

\draw[->] (R1S0) to (S0S0);
\draw[->] (R1S1) to (S0S1);

\draw[->] (R1R0) to (S1R0);
\draw[->] (R1R2) to (S1R1);

\draw[->] (R1S0) to (S1S0);
\draw[->] (R1S1) to (S1S0);

\draw[->] (R2R0) to (S1R0);
\draw[->] (R2R1) to (S1R1);

\draw[->] (R2S0) to (S1S0);
\end{tikzpicture}
	\caption{The layout of a 2-diagram.}
	\label{fig:layout}
\end{figure}

The layout algorithm is used to assign coordinates to every point in an $n$\-diagram, and is a crucial component of the rendering pipeline described in \Cref{subsec:rendering}, enabling rendering in 2D, 3D and 4D.
We use a categorical construction based on factorisation systems and colimits to extract a set of linear constraints from the total order data of the diagram~\cite{tataru2023layout}.
These constraints encode the necessary conditions for a layout to be well-defined.
For example, in 2D, this would include the information that one wire is to the left of another wire; or, in 3D, that one surface is in front of another.
We further impose aesthetic constraints, e.g.\ that wires and surfaces should be centred.
Finally, these constraints are passed to the linear solver HiGHS~\cite{huangfu2018parallelizing} to find a layout that satisfies all constraints.
An example layout for a 2-diagram is given in \Cref{fig:layout}.

\section{Implementation} \label{sec:implementation}

Here, we describe further implementation details: the memoisation techniques used in our core data structures and aspects of the rendering pipeline.
Whilst independent of the tool's mathematical foundations, we have found many of the details here essential, particularly in enabling the tool to operate in a resource-constrained environment---i.e., the browser.

\subsection{Memoisation}

Our data structures for \texttt{Diagram} and \texttt{Rewrite} are immutable.
Operations that would modify an object instead create a new one with the modifications applied.
This allows us to apply a technique known as \emph{hash consing}~\cite{filliatre2006type}: whenever a new \texttt{Diagram} or \texttt{Rewrite} is created, we check in a hash table if a logically equal instance is already present and reuse it if possible.
Hash consing enforces the invariant that any two instances that are logically equal must become physically equal---i.e., have the same representation and location in memory.
To test for equality we may therefore perform a simple pointer comparison instead of traversing the entire deeply nested structure; similarly, we cache the hash value of every \texttt{Diagram} and \texttt{Rewrite} to avoid deep traversals.
We incur a performance cost due to the hash table lookup involved whenever a new instance is created.
However, we observe significant improvements in performance and memory usage overall.
In practice, we have found that even under the sparse encoding of \Cref{sec:datastructures}, directly representing diagrams remains highly redundant: applying hash consing makes memory usage almost negligible, even for large diagrams.

The algorithms from \Cref{sec:algorithms} are largely recursive over the structure of a diagram and perform many repeated recursive calls on logically equal substructures.
We therefore memoise the results of recursive calls.
Since our deduplicated representation allows for very fast equality checks and caches hash values, lookups in the memoisation table are comparatively cheap.

Our data structures are reference counted, so we know when the last reference to an instance goes out of scope and it is safe to remove it from the hash table.
The recursive algorithms operating on diagrams temporarily materialise the structure that is implicit in the sparse encoding, leading to the same \texttt{Diagram} or \texttt{Rewrite} being created and destroyed many times during the execution of the algorithm.
An eager approach to maintaining the deduplication hash table therefore leads to unnecessary churn.
We therefore delay the removal of dead objects from the hash table to a batched garbage collection step that walks the table and removes instances that have no remaining references.

\subsection{Rendering pipeline} \label{subsec:rendering}

One of the most important parts of the implementation is the rendering pipeline, which allows for visualising $n$\-diagrams in up to four dimensions.
It consists of three components.
First, we have the \emph{layout} algorithm described in \Cref{subsec:layout} that assigns real coordinates to every point of an $n$\-diagram.
Second, there is the \emph{mesh generation}, which takes an $n$\-diagram and computes a cubical mesh---i.e., a subdivision of the $n$\-diagram into \emph{abstract} $k$\-dimensional cubes for $k \leq n$.
Together, these data are called a {\em geometry} which is a collection of $k$\-dimensional cuboids in the Euclidean space $\mathbb{R}^n$.
Finally, we have a \emph{subdivision} procedure which takes a geometry and makes it smoother by recursively subdividing each cube into smaller cubes with the positions of the new vertices calculated by interpolating the positions of the old vertices.

The subdivided geometry can then be sent to a number of rendering engines.
Amongst these, the most important are the SVG renderer, rendering 2D diagrams for all of the user's primary interaction, and the WebGL renderer, rendering 3D and 4D diagrams (with the 4D diagrams appearing as \emph{smooth} animations of 3D diagrams).
We also support a TikZ renderer, which can be used to generate high quality string diagrams for use in papers such as in this article.

In the 4-dimensional case, we obtain the animation by intersecting the geometry with an axis-aligned hyper-plane.
This computation is easily implemented via a compute shader.
However, to maximise \hio's compatibility with contemporary browsers, our shading pipeline is implemented in WebGL2, precluding the use of compute shaders.

\begin{figure}
	\centering
	\begin{tikzpicture}[scale=3, 3d view]
\coordinate (0) at (0, 0, 0);
\coordinate (1) at (1, 0, 0);
\coordinate (2) at (0, 1, 0);
\coordinate (3) at (1, 1, 0);
\coordinate (4) at (0, 0, 1);
\coordinate (5) at (1, 0, 1);
\coordinate (6) at (0, 1, 1);
\coordinate (7) at (1, 1, 1);

\draw[thick] (0) -- (1);
\draw[thick] (0) -- (2);
\draw[dotted] (0) -- (3);
\draw[thick] (0) -- (4);
\draw (0) -- (5);
\draw (0) -- (6);
\draw[dotted] (0) -- (7);

\draw[dashed] (1) -- (3);
\draw[thick] (1) -- (5);
\draw[dotted] (1) -- (7);

\draw[dashed] (2) -- (3);
\draw[thick] (2) -- (6);
\draw[dotted] (2) -- (7);

\draw[dashed] (3) -- (7);

\draw[thick] (4) -- (5);
\draw[thick] (4) -- (6);
\draw (4) -- (7);

\draw[thick] (5) -- (7);
\draw[thick] (6) -- (7);
\end{tikzpicture}
    \hspace{4em}
    \begin{tikzpicture}[scale=3, 3d view]
\coordinate (A) at (0, -0.8, 0);
\coordinate (B) at (1.5, 0.8, 0);
\coordinate (C) at (0, 1, 0);
\coordinate (D) at (0.4, 1, 0.9);

\draw[thick] (A) -- (B);
\draw[thick] (A) -- (C);
\draw[thick] (A) -- (D);
\draw[dashed] (B) -- (C);
\draw[thick] (B) -- (D);
\draw[thick] (C) -- (D);

\coordinate (BC) at ($ (C)!0.35!(B) $);
\coordinate (AC) at ($ (C)!0.35!(A) $);

\draw[dotted] (D) -- (BC);
\draw (D) -- (AC);
\draw[dotted] (BC) -- (AC);

\coordinate (BD) at ($ (D)!0.625!(B) $);
\coordinate (BC') at ($ (C)!0.8!(B) $);

\draw (A) -- (BD);
\draw[dotted] (A) -- (BC');
\draw[dotted] (BC') -- (BD);

\draw[dotted] (D) -- (BC');
\draw[dotted] (AC) -- (BC');

\end{tikzpicture}
	\caption{The decomposition of a 3-cube into six 3-simplices (left) and a 3-simplex into $w$-aligned sub-simplices (right).}
	\label{fig:cube-decomposition}
\end{figure}
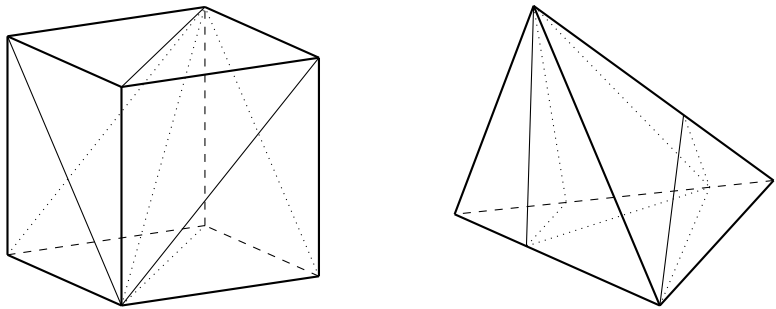

This situation presents a challenge, as computing intersection geometries offline for large diagrams is prohibitively expensive in terms of memory consumption.
Thus, we are forced perform the slicing in real-time, using a subtle rendering trick:
\begin{enumerate}
	\item each cube is broken down into a disjoint collection of covering simplices (\Cref{fig:cube-decomposition}, left);
	\item each simplex is further decomposed into sub-simplices, each with one face whose vertices have equal $w$-components,
	\item of the six edges of each simplex, the three with non-zero $w$-components are converted to three GL vertices, and the simplex a corresponding polygon (\Cref{fig:cube-decomposition}, right);
	\item each GL vertex is supplied with the start and end coordinates of the corresponding edge;
	\item the vertex shader interpolates between these points given a global $w$-coordinate, passed in as a uniform, dropping the vertex whenever $w$ is outside of the bounds of the edge.
\end{enumerate}
The effect is that the animation is broken down into a collection of \enquote{birth} and \enquote{death} events for small polygons, which can be rendered straightforwardly, as if it were a 3D mesh.
Animated wires in 4-dimensions are produced through a similar process, but the \enquote{pipe}-like appearance is produced via a deferred post-processing effect, to minimise the cost of storing geometry for the cylinders.

We additionally compute normals for intersection geometries in real-time via a barycentric interpolation across a volume-weighted average taken at each vertex.
This is directly analogous to the calculation of normals in traditional 3D pipelines.
However, much care is required when it comes to calculating and preserving the orientations of the elements of the mesh.

\section{Future work}\label{sec:future}

The tool is under active development, both mathematically as an end itself and as a piece of research software as a means to facilitate the development of higher-categorical mathematics.
We sketch some ideas for future improvements to the tool:
\begin{description}
    \item[Functor boxes] It would be both interesting and practical to consider what a functor would mean in this context; graphically, it should allow for \enquote{boxing} pieces of diagrams, allowing those components to be treated as atomic; this would present the user with more flexibility in representing proofs.
    \item[Dualisability support] The next logical step after the introduction of coherent inverses would be the introduction of coherent duals, which would allow for a mathematical exploration of important problems in higher algebra, such as the tangle hypothesis \cite{baez1995higher}, and knot theory.
    \item[Interopability with type theory] Some arguments of logical flavour are more easily expressed type-theoretically, e.g.\ in homotopy type theory or CaTT~\cite{finster2017type}, while more graphical arguments are better suited to \hio; ideally, we could make these tools talk to each other to support bimodal reasoning.
    \item[Linearity] Support for higher linear categories would make the tool amenable to directly performing calculations in topological quantum field theory.
    \item[Gallery] An arXiv-like gallery can be presented to showcase published proofs, better enabling proof discovery.
    \item[Cubical version] More general notions of string diagrams for cubical $n$-categories have been considered~\cite{myers2016string} and it would be an interesting question to extend \hio to cubical shapes.
\end{description}

\newpage
\bibliographystyle{plainurl}
\bibliography{bibliography}

\newpage

\appendix
\section{Case study: Hopf algebras and Hopf modules} \label{sec:casestudy}

In this appendix, we will detail an extended case study to highlight how \hio may be practically employed to formalise mathematical results of substantial character.
The basis we will build on for this case study will be the discussion in \Cref{sec:example}, showing that the Eckmann-Hilton argument could be used to formalise braids in the tool.
This will allow us to define Hopf algebras and Hopf modules in braided monoidal categories, and build towards a proof of the \emph{fundamental theorem of Hopf modules}, asserting the freeness of Hopf modules when certain idempotents split.
Since this formalisation will involve rich algebraic gadgets, long proofs and universal properties, we will build it up gradually.
The final formalisation, consisting of the full proofs, is available at \url{https://beta.homotopy.io/p/2402.00006} so that the reader can inspect the proofs in detail and visualise them in 3D and 4D.

Hopf algebras originated in work by Hopf~\cite{hopfuber1941} in topology, developed on and popularised by Milnor and Moore~\cite{milnorstructure1965} (see~\cite{andruskiewitschbeginnings2009} for more detail).
They have served to to generalise results from group theory to objects of study in topology, knot theory, algebraic geometry, combinatorics and quantum theory~\cite{sweedlerhopf1969}.
Our discussion here is informed by the original proof of the fundamental theorem of Hopf modules for Hopf algebras over principal ideal domains is given in~\cite{larsonassociative1969}, its generalisation to braided monoidal categories in~\cite{lyubashenkomodular1995} and the alternative proof using Karoubi completions given in~\cite{bespalov1998hopf}.
We also use the diagrammatic calculus for Hopf modules introduced by Majid in~\cite{majid1991braided}, which
has been used to give a string diagrammatic translation of this result in~\cite{zhang2005braided}, though the reader should be warned that the proof contains flaws.

\begin{figure}[b]
	\centering
	\input{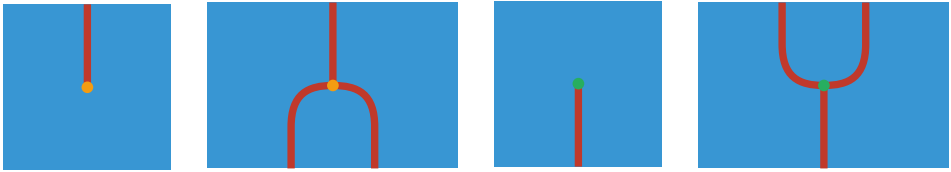}

\newcommand{\vvcenter}[1]{\begin{aligned}{\scalebox{0.55}{#1}}\end{aligned}}

\(
\vvcenter{
\begin{tikzpicture}
\begin{scope}
\fill[generator-0-0-1-pos] (0,0) -- (4,0) -- (4,4) -- (0,4) -- (0,0);
\draw[color=generator-1-2-0-pos, line width=5pt](2,2) -- (2,4);
\end{scope}
\fill[generator-2-3-0-pos] (2,2) circle (0.14);
\end{tikzpicture}
}
\quad
\vvcenter{
\begin{tikzpicture}
\begin{scope}
\fill[generator-0-0-1-pos] (0,0) -- (6,0) -- (6,4) -- (0,4) -- (0,0);
\draw[color=generator-1-2-0-pos, line width=5pt](2,0) -- (2,1) .. controls (2,1.8) and (2.4,2) .. (3,2) -- (3,4)(4,0) -- (4,1) .. controls (4,1.8) and (3.6,2) .. (3,2);
\end{scope}
\fill[generator-3-3-0-pos] (3,2) circle (0.14);
\end{tikzpicture}
}
\quad
\vvcenter{
\begin{tikzpicture}
\begin{scope}
\fill[generator-0-0-1-pos] (0,0) -- (4,0) -- (4,4) -- (0,4) -- (0,0);
\draw[color=generator-1-2-0-pos, line width=5pt](2,0) -- (2,2);
\end{scope}
\fill[generator-5-3-0-pos] (2,2) circle (0.14);
\end{tikzpicture}
}
\quad
\vvcenter{
\begin{tikzpicture}
\begin{scope}
\fill[generator-0-0-1-pos] (0,0) -- (6,0) -- (6,4) -- (0,4) -- (0,0);
\draw[color=generator-1-2-0-pos, line width=5pt](3,0) -- (3,2) .. controls (2.4,2) and (2,2.2) .. (2,3) -- (2,4)(3,2) .. controls (3.6,2) and (4,2.2) .. (4,3) -- (4,4);
\end{scope}
\fill[generator-4-3-0-pos] (3,2) circle (0.14);
\end{tikzpicture}
}
\)
	\caption{The 3-cells $\eta$, $\mu$, $\varepsilon$, and $\delta$ (left to right) in the bialgebra signature.}%
	\label{fig:bigalgsig}
\end{figure}

We start by first defining a bialgebra $H$ in a braided monoidal category.
Recall from \Cref{sec:example} that in order to obtain the braiding, we will start by adding a single 0-cell $x$ to our signature, and then add our 2-cell $H$ by specifying $1_x$ as source and target.
Hence $H$ is an object in our braided monoidal category.
Moreover, we need to add four 3-cells: a unit $\eta\colon 1 \to H$, a multiplication $\mu: H \otimes H \to H$, a counit $\varepsilon\colon H \to 1$, and a comultiplication $\delta: H \to H \otimes H$, which are depicted in \Cref{fig:bigalgsig}.

\begin{figure}
	\centering
	\input{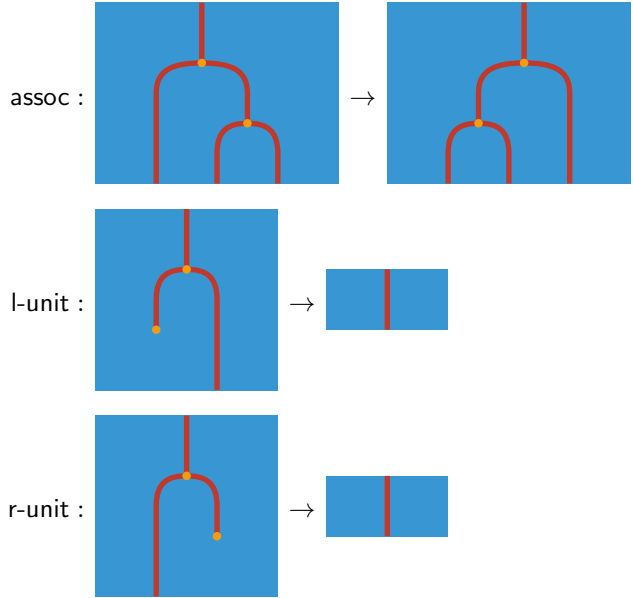}

\newcommand{\vvcenter}[1]{\begin{aligned}{\scalebox{0.4}{#1}}\end{aligned}}

\newcommand{\idH}{
\vvcenter{
\begin{tikzpicture}
\begin{scope}
\fill[generator-0-0-1-pos] (0,0) -- (4,0) -- (4,2) -- (0,2) -- (0,0);
\draw[color=generator-1-2-0-pos, line width=5pt](2,0) -- (2,2);
\end{scope}
\end{tikzpicture}
}
}

\begin{align*}
\textsf{assoc}
&:
\vvcenter{
\begin{tikzpicture}
\begin{scope}
\fill[generator-0-0-1-pos] (0,0) -- (8,0) -- (8,6) -- (0,6) -- (0,0);
\draw[color=generator-1-2-0-pos, line width=5pt](2,0) -- (2,3) .. controls (2,3.8) and (2.6,4) .. (3.5,4) -- (3.5,6)(4,0) -- (4,1) .. controls (4,1.8) and (4.4,2) .. (5,2) -- (5,3) .. controls (5,3.8) and (4.4,4) .. (3.5,4)(6,0) -- (6,1) .. controls (6,1.8) and (5.6,2) .. (5,2);
\end{scope}
\fill[generator-3-3-0-pos] (5,2) circle (0.14);
\fill[generator-3-3-0-pos] (3.5,4) circle (0.14);
\end{tikzpicture}
}
\to
\vvcenter{
\begin{tikzpicture}
\begin{scope}
\fill[generator-0-0-1-pos] (0,0) -- (8,0) -- (8,6) -- (0,6) -- (0,0);
\draw[color=generator-1-2-0-pos, line width=5pt](2,0) -- (2,1) .. controls (2,1.8) and (2.4,2) .. (3,2) -- (3,3) .. controls (3,3.8) and (3.6,4) .. (4.5,4) -- (4.5,6)(4,0) -- (4,1) .. controls (4,1.8) and (3.6,2) .. (3,2)(6,0) -- (6,3) .. controls (6,3.8) and (5.4,4) .. (4.5,4);
\end{scope}
\fill[generator-3-3-0-pos] (3,2) circle (0.14);
\fill[generator-3-3-0-pos] (4.5,4) circle (0.14);
\end{tikzpicture}
}
\\
\textsf{l-unit}
&:
\vvcenter{
\begin{tikzpicture}
\begin{scope}
\fill[generator-0-0-1-pos] (0,0) -- (6,0) -- (6,6) -- (0,6) -- (0,0);
\draw[color=generator-1-2-0-pos, line width=5pt](4,0) -- (4,3) .. controls (4,3.8) and (3.6,4) .. (3,4) -- (3,6)(2,2) -- (2,3) .. controls (2,3.8) and (2.4,4) .. (3,4);
\end{scope}
\fill[generator-2-3-0-pos] (2,2) circle (0.14);
\fill[generator-3-3-0-pos] (3,4) circle (0.14);
\end{tikzpicture}
}
\to
\idH
\\
\textsf{r-unit}
&:
\vvcenter{
\begin{tikzpicture}
\begin{scope}
\fill[generator-0-0-1-pos] (0,0) -- (6,0) -- (6,6) -- (0,6) -- (0,0);
\draw[color=generator-1-2-0-pos, line width=5pt](2,0) -- (2,3) .. controls (2,3.8) and (2.4,4) .. (3,4) -- (3,6)(4,2) -- (4,3) .. controls (4,3.8) and (3.6,4) .. (3,4);
\end{scope}
\fill[generator-2-3-0-pos] (4,2) circle (0.14);
\fill[generator-3-3-0-pos] (3,4) circle (0.14);
\end{tikzpicture}
}
\to
\idH
\end{align*}
	\caption{The associativity and unitality laws for the algebra $(H, \eta, \mu)$.}%
	\label{fig:alglaws}
\end{figure}

\begin{figure}
	\centering
	\input{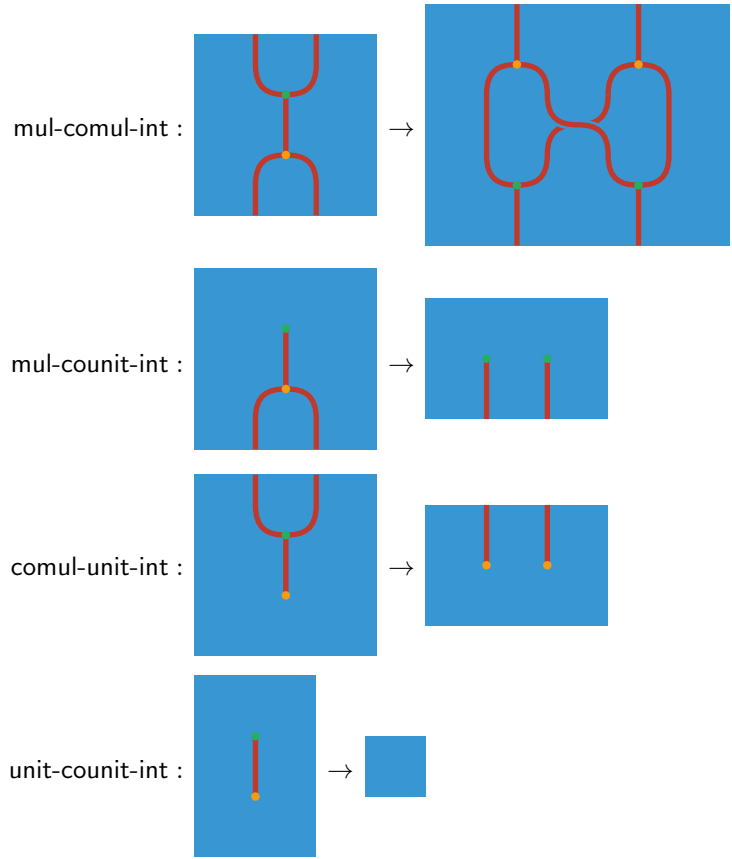}

\newcommand{\vvcenter}[1]{\begin{aligned}{\scalebox{0.4}{#1}}\end{aligned}}

\newcommand{\wire}[2]{
  \ifdefined\recolor\draw[color=\recolor, line width=10pt]\else\draw[color=#1, line width=5pt]\fi #2
}
\newcommand{\clipped}[3]{
\begin{scope}
  \newcommand{\recolor}{#1}
  \clip#3;
  #2
\end{scope}
}

\begin{align*}
\textsf{mul-comul-int}
&:
\vvcenter{
\begin{tikzpicture}
\begin{scope}
\fill[generator-0-0-1-pos] (0,0) -- (6,0) -- (6,6) -- (0,6) -- (0,0);
\draw[color=generator-1-2-0-pos, line width=5pt](2,0) -- (2,1) .. controls (2,1.8) and (2.4,2) .. (3,2) -- (3,4) .. controls (2.4,4) and (2,4.2) .. (2,5) -- (2,6)(4,0) -- (4,1) .. controls (4,1.8) and (3.6,2) .. (3,2)(3,4) .. controls (3.6,4) and (4,4.2) .. (4,5) -- (4,6);
\end{scope}
\fill[generator-3-3-0-pos] (3,2) circle (0.14);
\fill[generator-4-3-0-pos] (3,4) circle (0.14);
\end{tikzpicture}
}
\to
\vvcenter{
\begin{tikzpicture}
\begin{scope}[transparency group]
\fill[generator-0-0-1-pos] (0,0) -- (10,0) -- (10,8) -- (0,8) -- (0,0);
\newcommand{\layer}[1]{
  \clipped{generator-0-0-1-pos}{#1}{(0,0) -- (10,0) -- (10,8) -- (0,8) -- (0,0)}
  #1
}
\wire{generator-1-2-0-pos}{(4,3) .. controls (4,3.8) and (4.4,4) .. (5,4) .. controls (5.6,4) and (6,4.2) .. (6,5)};
\layer{
\wire{generator-1-2-0-pos}{(3,0) -- (3,2) .. controls (2.4,2) and (2,2.2) .. (2,3) -- (2,5) .. controls (2,5.8) and (2.4,6) .. (3,6) -- (3,8)(7,0) -- (7,2) .. controls (6.4,2) and (6,2.2) .. (6,3) .. controls (6,3.8) and (5.6,4) .. (5,4) .. controls (4.4,4) and (4,4.2) .. (4,5) .. controls (4,5.8) and (3.6,6) .. (3,6)(3,2) .. controls (3.6,2) and (4,2.2) .. (4,3)(7,2) .. controls (7.6,2) and (8,2.2) .. (8,3) -- (8,5) .. controls (8,5.8) and (7.6,6) .. (7,6) -- (7,8)(6,5) .. controls (6,5.8) and (6.4,6) .. (7,6)};
}
\end{scope}
\fill[generator-4-3-0-pos] (3,2) circle (0.14);
\fill[generator-4-3-0-pos] (7,2) circle (0.14);
\fill[generator-3-3-0-pos] (3,6) circle (0.14);
\fill[generator-3-3-0-pos] (7,6) circle (0.14);
\end{tikzpicture}
}
\\
\textsf{mul-counit-int}
&:
\vvcenter{
\begin{tikzpicture}
\begin{scope}
\fill[generator-0-0-1-pos] (0,0) -- (6,0) -- (6,6) -- (0,6) -- (0,0);
\draw[color=generator-1-2-0-pos, line width=5pt](2,0) -- (2,1) .. controls (2,1.8) and (2.4,2) .. (3,2) -- (3,4)(4,0) -- (4,1) .. controls (4,1.8) and (3.6,2) .. (3,2);
\end{scope}
\fill[generator-3-3-0-pos] (3,2) circle (0.14);
\fill[generator-5-3-0-pos] (3,4) circle (0.14);
\end{tikzpicture}
}
\to
\vvcenter{
\begin{tikzpicture}
\begin{scope}
\fill[generator-0-0-1-pos] (0,0) -- (6,0) -- (6,4) -- (0,4) -- (0,0);
\draw[color=generator-1-2-0-pos, line width=5pt](2,0) -- (2,2)(4,0) -- (4,2);
\end{scope}
\fill[generator-5-3-0-pos] (2,2) circle (0.14);
\fill[generator-5-3-0-pos] (4,2) circle (0.14);
\end{tikzpicture}
}
\\
\textsf{comul-unit-int}
&:
\vvcenter{
\begin{tikzpicture}
\begin{scope}
\fill[generator-0-0-1-pos] (0,0) -- (6,0) -- (6,6) -- (0,6) -- (0,0);
\draw[color=generator-1-2-0-pos, line width=5pt](3,2) -- (3,4) .. controls (2.4,4) and (2,4.2) .. (2,5) -- (2,6)(3,4) .. controls (3.6,4) and (4,4.2) .. (4,5) -- (4,6);
\end{scope}
\fill[generator-2-3-0-pos] (3,2) circle (0.14);
\fill[generator-4-3-0-pos] (3,4) circle (0.14);
\end{tikzpicture}
}
\to
\vvcenter{
\begin{tikzpicture}
\begin{scope}
\fill[generator-0-0-1-pos] (0,0) -- (6,0) -- (6,4) -- (0,4) -- (0,0);
\draw[color=generator-1-2-0-pos, line width=5pt](2,2) -- (2,4)(4,2) -- (4,4);
\end{scope}
\fill[generator-2-3-0-pos] (2,2) circle (0.14);
\fill[generator-2-3-0-pos] (4,2) circle (0.14);
\end{tikzpicture}
}
\\
\textsf{unit-counit-int}
&:
\vvcenter{
\begin{tikzpicture}
\begin{scope}
\fill[generator-0-0-1-pos] (0,0) -- (4,0) -- (4,6) -- (0,6) -- (0,0);
\draw[color=generator-1-2-0-pos, line width=5pt](2,2) -- (2,4);
\end{scope}
\fill[generator-2-3-0-pos] (2,2) circle (0.14);
\fill[generator-5-3-0-pos] (2,4) circle (0.14);
\end{tikzpicture}
}
\to
\vvcenter{
\begin{tikzpicture}
\begin{scope}
\fill[generator-0-0-1-pos] (0,0) -- (2,0) -- (2,2) -- (0,2) -- (0,0);
\end{scope}
\end{tikzpicture}
}
\end{align*}
	\caption{The bialgebra interaction laws for $(H, \eta, \mu, \varepsilon, \delta)$.}%
	\label{fig:bialglaws}
\end{figure}

For $H$ to be an \emph{algebra}, $\eta$ and $\mu$ must satisfy associativity and unitality laws, which are witnessed by the 4-cells depicted in \Cref{fig:alglaws}.
We add these to the signature and mark them as invertible.
For $H$ to be a \emph{coalgebra}, it will have to satisfy dual laws of coassociativity and counitality.
Finally, for $H$ to be a \emph{bialgebra}, the algebra and coalgebra structure have to interact: units must propagate through comultiplications, counits through multiplication, counit and units must be one-sided inverses, and multiplication and comultiplication have to interact via the braiding.
These interaction laws are specified as in \Cref{fig:bialglaws}.

\begin{figure}
	\centering
	\input{figures/hopf_stylesheet}

\newcommand{\vvcenter}[1]{\begin{aligned}{\scalebox{0.4}{#1}}\end{aligned}}

\begin{align*}
\textsf{l-antipode-elim}
&:
\vvcenter{
\begin{tikzpicture}
\begin{scope}
\fill[generator-0-0-1-pos] (0,0) -- (6,0) -- (6,8) -- (0,8) -- (0,0);
\draw[color=generator-1-2-0-pos, line width=5pt](3,0) -- (3,2) .. controls (2.4,2) and (2,2.2) .. (2,3) -- (2,5) .. controls (2,5.8) and (2.4,6) .. (3,6) -- (3,8)(3,2) .. controls (3.6,2) and (4,2.2) .. (4,3) -- (4,5) .. controls (4,5.8) and (3.6,6) .. (3,6);
\end{scope}
\fill[generator-4-3-0-pos] (3,2) circle (0.14);
\fill[generator-16-3-0-pos] (2,4) circle (0.14);
\fill[generator-3-3-0-pos] (3,6) circle (0.14);
\end{tikzpicture}
}
\to
\vvcenter{
\begin{tikzpicture}
\begin{scope}
\fill[generator-0-0-1-pos] (0,0) -- (4,0) -- (4,6) -- (0,6) -- (0,0);
\draw[color=generator-1-2-0-pos, line width=5pt](2,0) -- (2,2)(2,4) -- (2,6);
\end{scope}
\fill[generator-5-3-0-pos] (2,2) circle (0.14);
\fill[generator-2-3-0-pos] (2,4) circle (0.14);
\end{tikzpicture}
}
\\
\textsf{r-antipode-elim}
&:
\vvcenter{
\begin{tikzpicture}
\begin{scope}
\fill[generator-0-0-1-pos] (0,0) -- (6,0) -- (6,8) -- (0,8) -- (0,0);
\draw[color=generator-1-2-0-pos, line width=5pt](3,0) -- (3,2) .. controls (2.4,2) and (2,2.2) .. (2,3) -- (2,5) .. controls (2,5.8) and (2.4,6) .. (3,6) -- (3,8)(3,2) .. controls (3.6,2) and (4,2.2) .. (4,3) -- (4,5) .. controls (4,5.8) and (3.6,6) .. (3,6);
\end{scope}
\fill[generator-4-3-0-pos] (3,2) circle (0.14);
\fill[generator-16-3-0-pos] (4,4) circle (0.14);
\fill[generator-3-3-0-pos] (3,6) circle (0.14);
\end{tikzpicture}
}
\to
\vvcenter{
\begin{tikzpicture}
\begin{scope}
\fill[generator-0-0-1-pos] (0,0) -- (4,0) -- (4,6) -- (0,6) -- (0,0);
\draw[color=generator-1-2-0-pos, line width=5pt](2,0) -- (2,2)(2,4) -- (2,6);
\end{scope}
\fill[generator-5-3-0-pos] (2,2) circle (0.14);
\fill[generator-2-3-0-pos] (2,4) circle (0.14);
\end{tikzpicture}
}
\end{align*}
	\caption{The antipode cancellation laws for $\sigma\colon H \to H$.}%
	\label{fig:hopfantipodelaws}
\end{figure}

Thus $H$ is now a bialgebra. In order for it to be a \emph{Hopf algebra}, we must additionally equip it with an antipode map $\sigma\colon H \to H$.
This map needs to satisfy the two cancellation laws given in \Cref{fig:hopfantipodelaws}, which make it an inverse to the identity map on $H$ under the convolution product.
These two laws are enough to establish the following lemmas on the interactions between the antipode and the bialgebra structure.

\begin{lemma}
	\label{lemma:hopf_unitantipode}
	For a Hopf algebra $H$ in a braided monoidal category, we have $\sigma \circ \eta = \eta$.
\end{lemma}

\begin{figure}[h]
	\centering
	\input{figures/hopf_stylesheet}

\newcommand{\vvcenter}[1]{\begin{aligned}{\scalebox{0.4}{#1}}\end{aligned}}

\(
\vvcenter{
\begin{tikzpicture}
\begin{scope}
\fill[generator-0-0-1-pos] (0,0) -- (4,0) -- (4,6) -- (0,6) -- (0,0);
\draw[color=generator-1-2-0-pos, line width=5pt](2,2) -- (2,6);
\end{scope}
\fill[generator-2-3-0-pos] (2,2) circle (0.14);
\fill[generator-16-3-0-pos] (2,4) circle (0.14);
\end{tikzpicture}
}
\leadsto
\vvcenter{
\begin{tikzpicture}
\begin{scope}
\fill[generator-0-0-1-pos] (0,0) -- (6,0) -- (6,8) -- (0,8) -- (0,0);
\draw[color=generator-1-2-0-pos, line width=5pt](4,2) -- (4,5) .. controls (4,5.8) and (3.6,6) .. (3,6) -- (3,8)(2,4) -- (2,5) .. controls (2,5.8) and (2.4,6) .. (3,6);
\end{scope}
\fill[generator-2-3-0-pos] (4,2) circle (0.14);
\fill[generator-2-3-0-pos] (2,4) circle (0.14);
\fill[generator-16-3-0-pos] (4,4) circle (0.14);
\fill[generator-3-3-0-pos] (3,6) circle (0.14);
\end{tikzpicture}
}
\leadsto
\vvcenter{
\begin{tikzpicture}
\begin{scope}
\fill[generator-0-0-1-pos] (0,0) -- (6,0) -- (6,10) -- (0,10) -- (0,0);
\draw[color=generator-1-2-0-pos, line width=5pt](3,2) -- (3,4) .. controls (2.4,4) and (2,4.2) .. (2,5) -- (2,7) .. controls (2,7.8) and (2.4,8) .. (3,8) -- (3,10)(3,4) .. controls (3.6,4) and (4,4.2) .. (4,5) -- (4,7) .. controls (4,7.8) and (3.6,8) .. (3,8);
\end{scope}
\end{tikzpicture}
}
\leadsto
\vvcenter{
\begin{tikzpicture}
\begin{scope}
\fill[generator-0-0-1-pos] (0,0) -- (4,0) -- (4,8) -- (0,8) -- (0,0);
\draw[color=generator-1-2-0-pos, line width=5pt](2,2) -- (2,4)(2,6) -- (2,8);
\end{scope}
\fill[generator-2-3-0-pos] (2,2) circle (0.14);
\fill[generator-5-3-0-pos] (2,4) circle (0.14);
\fill[generator-2-3-0-pos] (2,6) circle (0.14);
\end{tikzpicture}
}
\leadsto
\vvcenter{
\begin{tikzpicture}
\begin{scope}
\fill[generator-0-0-1-pos] (0,0) -- (4,0) -- (4,4) -- (0,4) -- (0,0);
\draw[color=generator-1-2-0-pos, line width=5pt](2,2) -- (2,4);
\end{scope}
\fill[generator-2-3-0-pos] (2,2) circle (0.14);
\end{tikzpicture}
}
\)
	\caption{The proof of \Cref{lemma:hopf_unitantipode}.}
	\label{fig:hopf_unitantipode}
\end{figure}

\begin{lemma}[{\cite[Lemma 2.3]{majidalgebras1995}}]
	\label{lemma:hopf_mulantipode}
	For a Hopf algebra $H$ in a braided monoidal category, we have $(\sigma \otimes \sigma) \circ \gamma_{H,H} \circ \delta = \delta \circ \sigma$, where $\gamma_{H,H}\colon H \otimes H \to H \otimes H$ is the braiding.
\end{lemma}

\begin{proof}
	The essential steps of the graphical proof are given in \Cref{fig:hopfantipodelemma}.
	We start by constructing the left-hand side of the equation, as it involves a braid and thus is more complex than the right-hand side. This gradient of complexity helps us to use the contraction procedure effectively.

	The first step of the proof is to introduce units at the top of the diagram, and a counit at the bottom.
	We then braid the right-most unit past the antipode on the left, multiply the units
	and use them with the counit to introduce an antipode on the left.
	Above this antipode, we then use the multiplication/comultiplication axiom.
	Here we note that for proof to go through, the orientation of this braid, the braid in the starting diagram and the braid in the multiplication/comultiplication interaction axioms must match.

	A sequence of (co)associativity and braid naturality moves follows, until we can isolate the right-most antipode into a bubble to eliminate it into a unit/counit pair, which we use to remove a pair of (co)multiplications.
	We then move to use the (co)associativity laws to isolate the right-most antipode into a bubble, and proceed to simplify the remaining diagram down to our target diagram by cancelling the (co)units.
\end{proof}

At this point, our signature contains our Hopf algebra $H$ and two results about its properties, namely \Cref{lemma:hopf_unitantipode} and \Cref{lemma:hopf_mulantipode}.
Here we must note that the axioms for a Hopf algebra are self-dual, hence whenever we establish a result its dual statement, obtained by swapping unit and counit, multiplication and comultiplication and vice versa, will also hold. For instance, the dual of \Cref{lemma:hopf_unitantipode} would show that $\varepsilon \circ \sigma = \varepsilon$.

\begin{figure}
	\centering
	\input{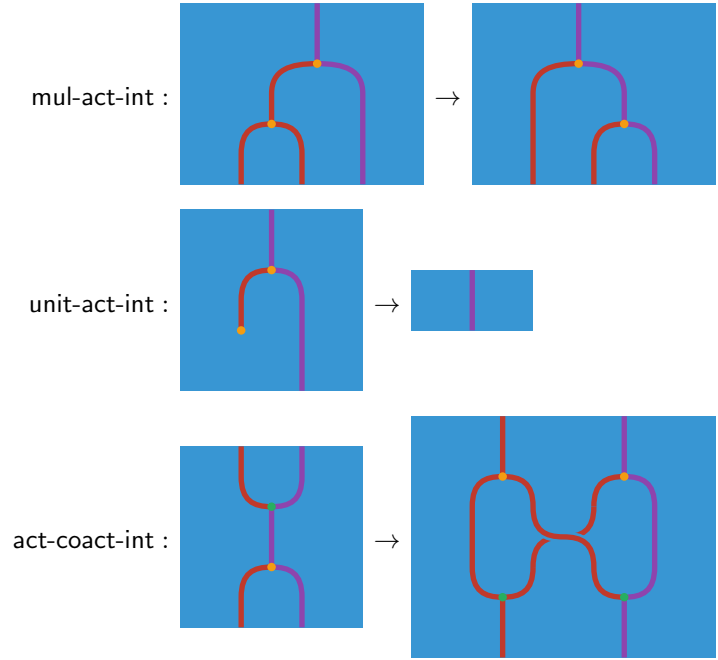}

\newcommand{\vvcenter}[1]{\begin{aligned}{\scalebox{0.4}{#1}}\end{aligned}}

\newcommand{\wire}[2]{
  \ifdefined\recolor\draw[color=\recolor, line width=10pt]\else\draw[color=#1, line width=5pt]\fi #2
}
\newcommand{\clipped}[3]{
\begin{scope}
  \newcommand{\recolor}{#1}
  \clip#3;
  #2
\end{scope}
}

\begin{align*}
\textsf{mul-act-int}
&:
\vvcenter{
\begin{tikzpicture}
\begin{scope}
\fill[generator-0-0-1-pos] (0,0) -- (8,0) -- (8,6) -- (0,6) -- (0,0);
\draw[color=generator-1-2-0-pos, line width=5pt](2,0) -- (2,1) .. controls (2,1.8) and (2.4,2) .. (3,2) -- (3,3) .. controls (3,3.8) and (3.6,4) .. (4.5,4)(4,0) -- (4,1) .. controls (4,1.8) and (3.6,2) .. (3,2);
\draw[color=generator-19-2-0-pos, line width=5pt](6,0) -- (6,3) .. controls (6,3.8) and (5.4,4) .. (4.5,4) -- (4.5,6);
\end{scope}
\fill[generator-3-3-0-pos] (3,2) circle (0.14);
\fill[generator-20-3-0-pos] (4.5,4) circle (0.14);
\end{tikzpicture}
}
\to
\vvcenter{
\begin{tikzpicture}
\begin{scope}
\fill[generator-0-0-1-pos] (0,0) -- (8,0) -- (8,6) -- (0,6) -- (0,0);
\draw[color=generator-1-2-0-pos, line width=5pt](2,0) -- (2,3) .. controls (2,3.8) and (2.6,4) .. (3.5,4)(4,0) -- (4,1) .. controls (4,1.8) and (4.4,2) .. (5,2);
\draw[color=generator-19-2-0-pos, line width=5pt](6,0) -- (6,1) .. controls (6,1.8) and (5.6,2) .. (5,2) -- (5,3) .. controls (5,3.8) and (4.4,4) .. (3.5,4) -- (3.5,6);
\end{scope}
\fill[generator-20-3-0-pos] (5,2) circle (0.14);
\fill[generator-20-3-0-pos] (3.5,4) circle (0.14);
\end{tikzpicture}
}
\\
\textsf{unit-act-int}
&:
\vvcenter{
\begin{tikzpicture}
\begin{scope}
\fill[generator-0-0-1-pos] (0,0) -- (6,0) -- (6,6) -- (0,6) -- (0,0);
\draw[color=generator-1-2-0-pos, line width=5pt](2,2) -- (2,3) .. controls (2,3.8) and (2.4,4) .. (3,4);
\draw[color=generator-19-2-0-pos, line width=5pt](4,0) -- (4,3) .. controls (4,3.8) and (3.6,4) .. (3,4) -- (3,6);
\end{scope}
\fill[generator-2-3-0-pos] (2,2) circle (0.14);
\fill[generator-20-3-0-pos] (3,4) circle (0.14);
\end{tikzpicture}
}
\to
\vvcenter{
\begin{tikzpicture}
\begin{scope}
\fill[generator-0-0-1-pos] (0,0) -- (4,0) -- (4,2) -- (0,2) -- (0,0);
\draw[color=generator-19-2-0-pos, line width=5pt](2,0) -- (2,2);
\end{scope}
\end{tikzpicture}
}
\\
\textsf{act-coact-int}
&:
\vvcenter{
\begin{tikzpicture}
\begin{scope}
\fill[generator-0-0-1-pos] (0,0) -- (6,0) -- (6,6) -- (0,6) -- (0,0);
\draw[color=generator-1-2-0-pos, line width=5pt](2,0) -- (2,1) .. controls (2,1.8) and (2.4,2) .. (3,2)(3,4) .. controls (2.4,4) and (2,4.2) .. (2,5) -- (2,6);
\draw[color=generator-19-2-0-pos, line width=5pt](4,0) -- (4,1) .. controls (4,1.8) and (3.6,2) .. (3,2) -- (3,4) .. controls (3.6,4) and (4,4.2) .. (4,5) -- (4,6);
\end{scope}
\fill[generator-20-3-0-pos] (3,2) circle (0.14);
\fill[generator-21-3-0-pos] (3,4) circle (0.14);
\end{tikzpicture}
}
\to
\vvcenter{
\begin{tikzpicture}
\begin{scope}[transparency group]
\fill[generator-0-0-1-pos] (0,0) -- (10,0) -- (10,8) -- (0,8) -- (0,0);
\newcommand{\layer}[1]{
  \clipped{generator-0-0-1-pos}{#1}{(0,0) -- (10,0) -- (10,8) -- (0,8) -- (0,0)}
  #1
}
\wire{generator-1-2-0-pos}{(4,3) .. controls (4,3.8) and (4.4,4) .. (5,4) .. controls (5.6,4) and (6,4.2) .. (6,5)};
\layer{
\wire{generator-1-2-0-pos}{(3,0) -- (3,2) .. controls (2.4,2) and (2,2.2) .. (2,3) -- (2,5) .. controls (2,5.8) and (2.4,6) .. (3,6) -- (3,8)(3,2) .. controls (3.6,2) and (4,2.2) .. (4,3)(7,2) .. controls (6.4,2) and (6,2.2) .. (6,3) .. controls (6,3.8) and (5.6,4) .. (5,4) .. controls (4.4,4) and (4,4.2) .. (4,5) .. controls (4,5.8) and (3.6,6) .. (3,6)(6,5) .. controls (6,5.8) and (6.4,6) .. (7,6)};
\wire{generator-19-2-0-pos}{(7,0) -- (7,2) .. controls (7.6,2) and (8,2.2) .. (8,3) -- (8,5) .. controls (8,5.8) and (7.6,6) .. (7,6) -- (7,8)};
}
\end{scope}
\fill[generator-4-3-0-pos] (3,2) circle (0.14);
\fill[generator-21-3-0-pos] (7,2) circle (0.14);
\fill[generator-3-3-0-pos] (3,6) circle (0.14);
\fill[generator-20-3-0-pos] (7,6) circle (0.14);
\end{tikzpicture}
}
\end{align*}
	\caption{The $H$-Hopf module laws for $M$, with the duals to the first two omitted.}
	\label{fig:hopfmodulelaws}
\end{figure}

We will now add another object $M$ with the same boundary as $H$, which is to be a \emph{(left) $H$-Hopf module}.
This means $M$ is equipped with two maps, a (left) action $\alpha\colon H \otimes M \to M$ and a (left) coaction $\phi\colon M \to H \otimes M$, which are compatible with the structure of $H$.
Thus we add invertible cells as in \Cref{fig:hopfmodulelaws} to impose interaction between the multiplication $\mu$ and the action $\alpha$, between the unit $\eta$ and $\alpha$, their duals for $\delta,\varepsilon$ and $\phi$, and finally the interaction between the action/coaction pair.

We now have an $H$-Hopf module $M$ which we have manually added to the signature.
The structure of the Hopf algebra $H$ gives us another way to obtain $H$-Hopf modules: given any any object $B$ in our braided monoidal category, we may equip $H \otimes B$ with the structure of an $H$-Hopf module by taking the action to be $\mu \otimes 1_B\colon H \otimes H \otimes B \to H \otimes B$ and the coaction to be $\delta \otimes 1_B\colon H \otimes B \to H \otimes H \otimes B$.
We call Hopf modules of this form \emph{free}.

\begin{figure}
	\centering
	\input{figures/hopf_stylesheet}

\scalebox{0.4}{
\begin{tikzpicture}
\begin{scope}
\fill[generator-0-0-1-pos] (0,0) -- (6,0) -- (6,8) -- (0,8) -- (0,0);
\draw[color=generator-1-2-0-pos, line width=5pt](3,2) .. controls (2.4,2) and (2,2.2) .. (2,3) -- (2,5) .. controls (2,5.8) and (2.4,6) .. (3,6);
\draw[color=generator-19-2-0-pos, line width=5pt](3,0) -- (3,2) .. controls (3.6,2) and (4,2.2) .. (4,3) -- (4,5) .. controls (4,5.8) and (3.6,6) .. (3,6) -- (3,8);
\end{scope}
\fill[generator-21-3-0-pos] (3,2) circle (0.14);
\fill[generator-16-3-0-pos] (2,4) circle (0.14);
\fill[generator-20-3-0-pos] (3,6) circle (0.14);
\end{tikzpicture}
}
	\caption{The idempotent endomorphism $\nu\colon M \to M$.}
	\label{fig:coninvmap}
\end{figure}

With the action and coaction of $M$ and the antipode of $H$, we may define the endomorphism $\nu \coloneqq \alpha \circ (\sigma \otimes 1_M) \circ \phi\colon M \to M$ depicted in \Cref{fig:coninvmap}.
It turns out that this map is idempotent.

\begin{lemma}[{\cite[Prop.3.2.1]{bespalov1998hopf}}]
	\label{lemma:hopf_idempotency}
	The map $\nu\colon M \to M$ is idempotent---i.e., $\nu \circ \nu = \nu$.
\end{lemma}

\begin{figure}
	\centering
	\input{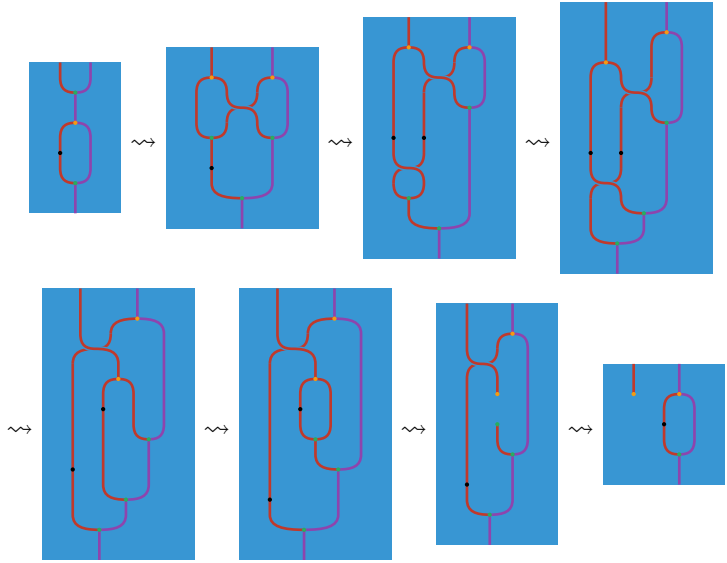}
	\caption{The proof of $\phi \circ \nu = (\eta \otimes 1_M) \circ \nu$ in \Cref{lemma:hopf_idempotency}.}%
	\label{fig:hopfidemplemma}
\end{figure}

\begin{proof}
	We first use \Cref{lemma:hopf_mulantipode} to establish that $\phi \circ \nu = (\eta \otimes 1_M) \circ \nu$.
	This is given in \Cref{fig:hopfidemplemma}, where the key step is to use the action/coaction interaction axiom in order to apply \Cref{lemma:hopf_idempotency}, and then simplify down the diagram.
	With this at hand, idempotency follows from \Cref{lemma:hopf_unitantipode} and routine simplifications.
\end{proof}

With \Cref{lemma:hopf_idempotency} at hand, we assume now that we have a splitting for $\nu$---i.e., an object $\hM$ and maps $\iota\colon \hM \to M$, $\upsilon\colon M \to \hM$ such that $\upsilon\circ \iota = 1_{\hM}$ and $\iota \circ \upsilon = \nu$.
It turns out that this splitting enjoys an additional universal property.

\begin{lemma}[{\cite[Prop.3.2.1]{bespalov1998hopf}}]
	\label{lemma:hopf_split_eq}
	The map $\iota\colon \hM \to M$ is the equaliser of $\phi$ and $\eta \otimes 1_M$.
\end{lemma}

\begin{proof}
	We add an object $P$ and a map $\chi\colon P \to M$ such that $\phi \circ \chi = \eta \otimes \chi$ to the signature.
	We then show that $\chi = \nu \circ \chi$ by applying \Cref{lemma:hopf_unitantipode}.
	But this means $\chi = \iota \circ (\upsilon \circ \chi)$ and since $\iota$ is monic this factorisation must be unique.
\end{proof}

Thus $\hM$ enjoys a universal property which trivialises the coaction, we call it \emph{object of coinvariants} of $M$.
Note that by dualising \Cref{lemma:hopf_split_eq} and its dependencies, we may also prove that $\upsilon$ is the coequaliser of $\alpha$ and $\varepsilon \otimes 1_M$, and thus $\hM$ is also the \emph{object of invariants} of $M$.
Now the fundamental theorem of Hopf modules then asserts that if idempotents split in a braided monoidal category, up to isomorphism, every Hopf module is free.

\begin{theorem}[Fundamental Theorem of Hopf Modules {\cite[Prop. 1]{larsonassociative1969},\cite[Lemma 3.3.3]{bespalov1998hopf}}]
	\label{thm:hopf_fundmod}
	Let $M$ be a left $H$-Hopf module in a braided monoidal category, and assume the map $\nu\colon M \to M$ splits through $\hM$.
	There is an isomorphism of left $H$-Hopf modules $H \otimes \hM \cong M$.
\end{theorem}

\begin{proof}
	We use the data of the splitting $\nu = \iota \circ \upsilon$ to construct two comparison maps $\xi \coloneqq \alpha \circ (1_H \otimes \iota)$ and $\psi \coloneqq (1_H \otimes \upsilon) \circ \phi$ between $H \otimes \hM$ and $M$, which will show induce the isomorphism of Hopf modules.
	The string diagrams corresponding to $\xi$ and $\psi$ are given in \Cref{fig:hopf_comparison}.
	\begin{figure}[h]
		\centering
		\begin{subfigure}{0.49\textwidth}
			\centering
			\scalebox{0.5}{\begin{tikzpicture}
\input{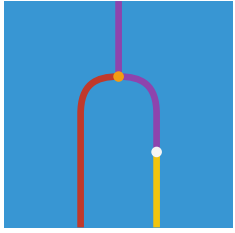}
\begin{scope}
\fill[generator-0-0-1-pos] (0,0) -- (6,0) -- (6,6) -- (0,6) -- (0,0);
\draw[color=generator-19-2-0-pos, line width=5pt](4,2) -- (4,3) .. controls (4,3.8) and (3.6,4) .. (3,4) -- (3,6);
\draw[color=generator-1-2-0-pos, line width=5pt](2,0) -- (2,3) .. controls (2,3.8) and (2.4,4) .. (3,4);
\draw[color=generator-29-2-0-pos, line width=5pt](4,0) -- (4,2);
\end{scope}
\fill[generator-30-3-0-pos] (4,2) circle (0.14);
\fill[generator-20-3-0-pos] (3,4) circle (0.14);
\end{tikzpicture}}
			\caption{The map $\xi\colon H \otimes \hM \to M$.}
		\end{subfigure}
		\begin{subfigure}{0.49\textwidth}
			\centering
			\scalebox{0.5}{\begin{tikzpicture}
\input{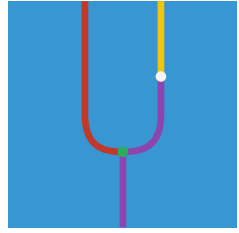}

\begin{scope}
\fill[generator-0-0-1-pos] (0,0) -- (6,0) -- (6,6) -- (0,6) -- (0,0);
\draw[color=generator-29-2-0-pos, line width=5pt](4,4) -- (4,6);
\draw[color=generator-1-2-0-pos, line width=5pt](3,2) .. controls (2.4,2) and (2,2.2) .. (2,3) -- (2,6);
\draw[color=generator-19-2-0-pos, line width=5pt](3,0) -- (3,2) .. controls (3.6,2) and (4,2.2) .. (4,3) -- (4,4);
\end{scope}
\fill[generator-21-3-0-pos] (3,2) circle (0.14);
\fill[generator-36-3-0-pos] (4,4) circle (0.14);
\end{tikzpicture}}
			\caption{The map $\psi\colon M \to H \otimes \hM$.}
		\end{subfigure}
		\caption{The string diagrams for the mutually inverse $\xi$ and $\psi$.}
		\label{fig:hopf_comparison}
	\end{figure}

	We claim $\xi$ respects the Hopf module structure. Compatibility between $\xi$ and the actions $\alpha$ and $\mu \otimes 1_{\hM}$ follows immediately from the multiplication/action interaction law.
	As for the coactions, we follow \Cref{fig:hopf_compcoact} in using the action/coaction interaction law, followed by the equaliser property given by \Cref{lemma:hopf_split_eq}. This establishes the claim.
	\begin{figure}
		\centering
		\input{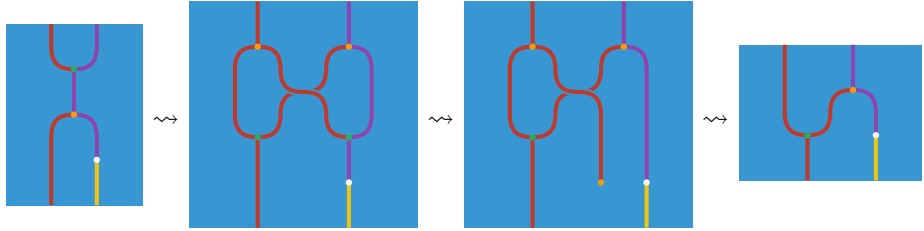}

\newcommand{\vvcenter}[1]{\begin{aligned}{\scalebox{0.3}{#1}}\end{aligned}}

\newcommand{\wire}[2]{
  \ifdefined\recolor\draw[color=\recolor, line width=10pt]\else\draw[color=#1, line width=5pt]\fi #2
}
\newcommand{\clipped}[3]{
\begin{scope}
  \newcommand{\recolor}{#1}
  \clip#3;
  #2
\end{scope}
}

\(
\vvcenter{
\begin{tikzpicture}
\begin{scope}
\fill[generator-0-0-1-pos] (0,0) -- (6,0) -- (6,8) -- (0,8) -- (0,0);
\draw[color=generator-19-2-0-pos, line width=5pt](4,2) -- (4,3) .. controls (4,3.8) and (3.6,4) .. (3,4) -- (3,6) .. controls (3.6,6) and (4,6.2) .. (4,7) -- (4,8);
\draw[color=generator-1-2-0-pos, line width=5pt](2,0) -- (2,3) .. controls (2,3.8) and (2.4,4) .. (3,4)(3,6) .. controls (2.4,6) and (2,6.2) .. (2,7) -- (2,8);
\draw[color=generator-29-2-0-pos, line width=5pt](4,0) -- (4,2);
\end{scope}
\fill[generator-30-3-0-pos] (4,2) circle (0.14);
\fill[generator-20-3-0-pos] (3,4) circle (0.14);
\fill[generator-21-3-0-pos] (3,6) circle (0.14);
\end{tikzpicture}
}
\leadsto
\vvcenter{
\begin{tikzpicture}
\begin{scope}[transparency group]
\fill[generator-0-0-1-pos] (0,0) -- (10,0) -- (10,10) -- (0,10) -- (0,0);
\newcommand{\layer}[1]{
  \clipped{generator-0-0-1-pos}{#1}{(0,0) -- (10,0) -- (10,10) -- (0,10) -- (0,0)}
  #1
}

\wire{generator-1-2-0-pos}{(4,5) .. controls (4,5.8) and (4.4,6) .. (5,6) .. controls (5.6,6) and (6,6.2) .. (6,7)};
\layer{
\wire{generator-1-2-0-pos}{(3,0) -- (3,4) .. controls (2.4,4) and (2,4.2) .. (2,5) -- (2,7) .. controls (2,7.8) and (2.4,8) .. (3,8) -- (3,10)(3,4) .. controls (3.6,4) and (4,4.2) .. (4,5)(7,4) .. controls (6.4,4) and (6,4.2) .. (6,5) .. controls (6,5.8) and (5.6,6) .. (5,6) .. controls (4.4,6) and (4,6.2) .. (4,7) .. controls (4,7.8) and (3.6,8) .. (3,8)(6,7) .. controls (6,7.8) and (6.4,8) .. (7,8)};
\wire{generator-19-2-0-pos}{(7,2) -- (7,4) .. controls (7.6,4) and (8,4.2) .. (8,5) -- (8,7) .. controls (8,7.8) and (7.6,8) .. (7,8) -- (7,10)};
\wire{generator-29-2-0-pos}{(7,0) -- (7,2)};
}
\end{scope}
\fill[generator-30-3-0-pos] (7,2) circle (0.14);
\fill[generator-4-3-0-pos] (3,4) circle (0.14);
\fill[generator-21-3-0-pos] (7,4) circle (0.14);
\fill[generator-3-3-0-pos] (3,8) circle (0.14);
\fill[generator-20-3-0-pos] (7,8) circle (0.14);
\end{tikzpicture}
}
\leadsto
\vvcenter{
\begin{tikzpicture}
\begin{scope}[transparency group]
\fill[generator-0-0-1-pos] (0,0) -- (10,0) -- (10,10) -- (0,10) -- (0,0);
\newcommand{\layer}[1]{
  \clipped{generator-0-0-1-pos}{#1}{(0,0) -- (10,0) -- (10,10) -- (0,10) -- (0,0)}
  #1
}

\wire{generator-1-2-0-pos}{(4,5) .. controls (4,5.8) and (4.4,6) .. (5,6) .. controls (5.6,6) and (6,6.2) .. (6,7)};
\layer{
\wire{generator-1-2-0-pos}{(3,0) -- (3,4) .. controls (2.4,4) and (2,4.2) .. (2,5) -- (2,7) .. controls (2,7.8) and (2.4,8) .. (3,8) -- (3,10)(6,2) -- (6,5) .. controls (6,5.8) and (5.6,6) .. (5,6) .. controls (4.4,6) and (4,6.2) .. (4,7) .. controls (4,7.8) and (3.6,8) .. (3,8)(3,4) .. controls (3.6,4) and (4,4.2) .. (4,5)(6,7) .. controls (6,7.8) and (6.4,8) .. (7,8)};
\wire{generator-19-2-0-pos}{(8,2) -- (8,7) .. controls (8,7.8) and (7.6,8) .. (7,8) -- (7,10)};
\wire{generator-29-2-0-pos}{(8,0) -- (8,2)};
}
\end{scope}
\fill[generator-2-3-0-pos] (6,2) circle (0.14);
\fill[generator-30-3-0-pos] (8,2) circle (0.14);
\fill[generator-4-3-0-pos] (3,4) circle (0.14);
\fill[generator-3-3-0-pos] (3,8) circle (0.14);
\fill[generator-20-3-0-pos] (7,8) circle (0.14);
\end{tikzpicture}
}
\leadsto
\vvcenter{
\begin{tikzpicture}
\begin{tikzpicture}
\begin{scope}
\fill[generator-0-0-1-pos] (0,0) -- (8,0) -- (8,6) -- (0,6) -- (0,0);
\draw[color=generator-19-2-0-pos, line width=5pt](6,2) -- (6,3) .. controls (6,3.8) and (5.6,4) .. (5,4) -- (5,6);
\draw[color=generator-1-2-0-pos, line width=5pt](3,0) -- (3,2) .. controls (2.4,2) and (2,2.2) .. (2,3) -- (2,6)(3,2) .. controls (3.6,2) and (4,2.2) .. (4,3) .. controls (4,3.8) and (4.4,4) .. (5,4);
\draw[color=generator-29-2-0-pos, line width=5pt](6,0) -- (6,2);
\end{scope}
\fill[generator-4-3-0-pos] (3,2) circle (0.14);
\fill[generator-30-3-0-pos] (6,2) circle (0.14);
\fill[generator-20-3-0-pos] (5,4) circle (0.14);
\end{tikzpicture}
\end{tikzpicture}
}
\)
		\caption{The map $\xi$ respects the coactions $\phi$ and $\delta \otimes 1_{\hM}$.}
		\label{fig:hopf_compcoact}
	\end{figure}

\begin{figure}
	\centering
	\input{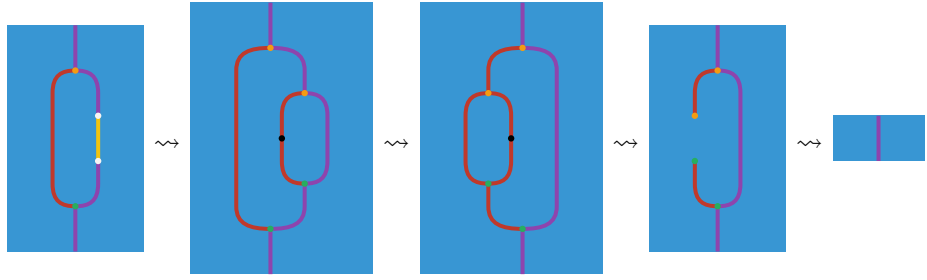}

\newcommand{\vvcenter}[1]{\begin{aligned}{\scalebox{0.3}{#1}}\end{aligned}}

\(
\vvcenter{
\begin{tikzpicture}
\begin{scope}
\fill[generator-0-0-1-pos] (0,0) -- (6,0) -- (6,10) -- (0,10) -- (0,0);
\draw[color=generator-29-2-0-pos, line width=5pt](4,4) -- (4,6);
\draw[color=generator-1-2-0-pos, line width=5pt](3,2) .. controls (2.4,2) and (2,2.2) .. (2,3) -- (2,7) .. controls (2,7.8) and (2.4,8) .. (3,8);
\draw[color=generator-19-2-0-pos, line width=5pt](3,0) -- (3,2) .. controls (3.6,2) and (4,2.2) .. (4,3) -- (4,4)(4,6) -- (4,7) .. controls (4,7.8) and (3.6,8) .. (3,8) -- (3,10);
\end{scope}
\fill[generator-21-3-0-pos] (3,2) circle (0.14);
\fill[generator-36-3-0-pos] (4,4) circle (0.14);
\fill[generator-30-3-0-pos] (4,6) circle (0.14);
\fill[generator-20-3-0-pos] (3,8) circle (0.14);
\end{tikzpicture}
}
\leadsto
\vvcenter{
\begin{tikzpicture}
\begin{scope}
\fill[generator-0-0-1-pos] (0,0) -- (8,0) -- (8,12) -- (0,12) -- (0,0);
\draw[color=generator-1-2-0-pos, line width=5pt](3.5,2) .. controls (2.6,2) and (2,2.2) .. (2,3) -- (2,9) .. controls (2,9.8) and (2.6,10) .. (3.5,10)(5,4) .. controls (4.4,4) and (4,4.2) .. (4,5) -- (4,7) .. controls (4,7.8) and (4.4,8) .. (5,8);
\draw[color=generator-19-2-0-pos, line width=5pt](3.5,0) -- (3.5,2) .. controls (4.4,2) and (5,2.2) .. (5,3) -- (5,4) .. controls (5.6,4) and (6,4.2) .. (6,5) -- (6,7) .. controls (6,7.8) and (5.6,8) .. (5,8) -- (5,9) .. controls (5,9.8) and (4.4,10) .. (3.5,10) -- (3.5,12);
\end{scope}
\fill[generator-21-3-0-pos] (3.5,2) circle (0.14);
\fill[generator-21-3-0-pos] (5,4) circle (0.14);
\fill[generator-16-3-0-pos] (4,6) circle (0.14);
\fill[generator-20-3-0-pos] (5,8) circle (0.14);
\fill[generator-20-3-0-pos] (3.5,10) circle (0.14);
\end{tikzpicture}
}
\leadsto
\vvcenter{
\begin{tikzpicture}
\begin{scope}
\fill[generator-0-0-1-pos] (0,0) -- (8,0) -- (8,12) -- (0,12) -- (0,0);
\draw[color=generator-1-2-0-pos, line width=5pt](4.5,2) .. controls (3.6,2) and (3,2.2) .. (3,3) -- (3,4) .. controls (2.4,4) and (2,4.2) .. (2,5) -- (2,7) .. controls (2,7.8) and (2.4,8) .. (3,8) -- (3,9) .. controls (3,9.8) and (3.6,10) .. (4.5,10)(3,4) .. controls (3.6,4) and (4,4.2) .. (4,5) -- (4,7) .. controls (4,7.8) and (3.6,8) .. (3,8);
\draw[color=generator-19-2-0-pos, line width=5pt](4.5,0) -- (4.5,2) .. controls (5.4,2) and (6,2.2) .. (6,3) -- (6,9) .. controls (6,9.8) and (5.4,10) .. (4.5,10) -- (4.5,12);
\end{scope}
\fill[generator-21-3-0-pos] (4.5,2) circle (0.14);
\fill[generator-4-3-0-pos] (3,4) circle (0.14);
\fill[generator-16-3-0-pos] (4,6) circle (0.14);
\fill[generator-3-3-0-pos] (3,8) circle (0.14);
\fill[generator-20-3-0-pos] (4.5,10) circle (0.14);
\end{tikzpicture}
}
\leadsto
\vvcenter{
\begin{tikzpicture}
\begin{scope}
\fill[generator-0-0-1-pos] (0,0) -- (6,0) -- (6,10) -- (0,10) -- (0,0);
\draw[color=generator-1-2-0-pos, line width=5pt](3,2) .. controls (2.4,2) and (2,2.2) .. (2,3) -- (2,4)(2,6) -- (2,7) .. controls (2,7.8) and (2.4,8) .. (3,8);
\draw[color=generator-19-2-0-pos, line width=5pt](3,0) -- (3,2) .. controls (3.6,2) and (4,2.2) .. (4,3) -- (4,7) .. controls (4,7.8) and (3.6,8) .. (3,8) -- (3,10);
\end{scope}
\fill[generator-21-3-0-pos] (3,2) circle (0.14);
\fill[generator-5-3-0-pos] (2,4) circle (0.14);
\fill[generator-2-3-0-pos] (2,6) circle (0.14);
\fill[generator-20-3-0-pos] (3,8) circle (0.14);
\end{tikzpicture}
}
\leadsto
\vvcenter{
\begin{tikzpicture}
\begin{scope}
\fill[generator-0-0-1-pos] (0,0) -- (4,0) -- (4,2) -- (0,2) -- (0,0);
\draw[color=generator-19-2-0-pos, line width=5pt](2,0) -- (2,2);
\end{scope}
\end{tikzpicture}
}
\)
	\caption{The proof of $\xi \circ \psi = 1_M$ in \Cref{thm:hopf_fundmod}.}
	\label{fig:hopf_iso1}
\end{figure}

\begin{figure}
	\centering
	\input{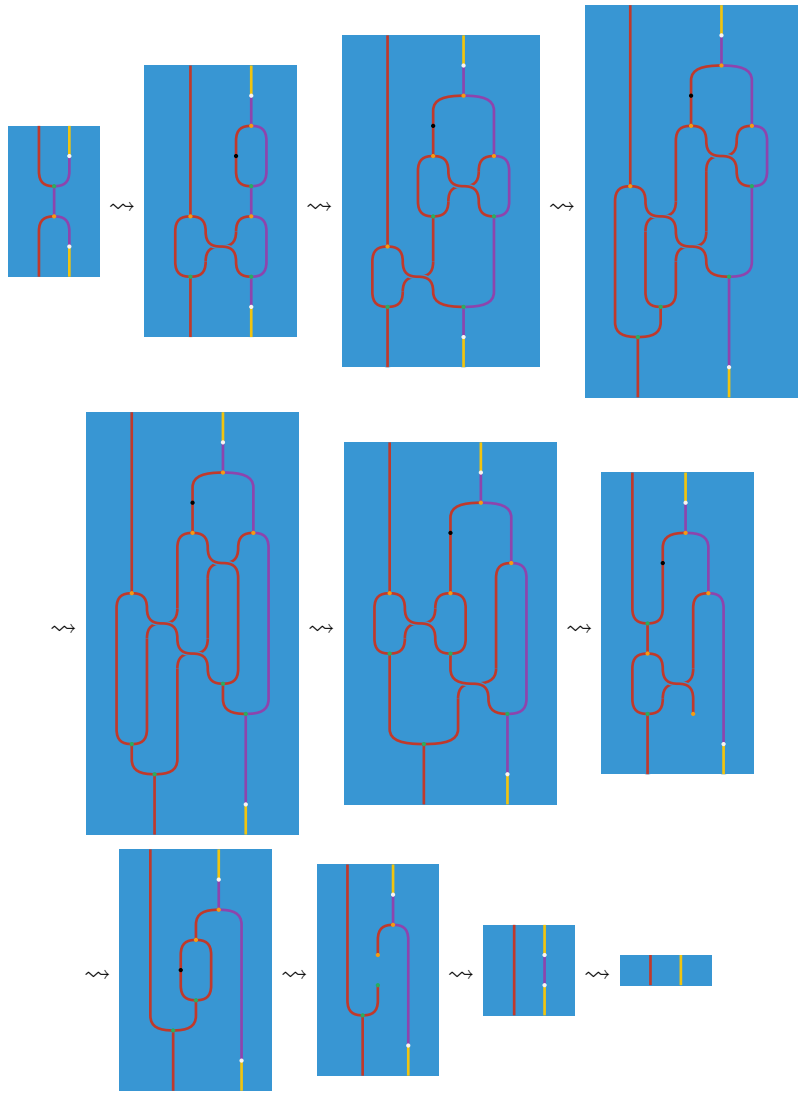}
	\caption{The proof of $\psi \circ \xi = 1_{H \otimes \hM}$ in \Cref{thm:hopf_fundmod}.}
	\label{fig:hopf_iso2}
\end{figure}

	Finally, we prove that $\xi$ and $\psi$ are mutual inverses.
	The proof of $\xi \circ \psi = 1_M$ is simpler, consisting mainly in the cancellation of the antipode, and is given in \Cref{fig:hopf_iso1}.
	The proof of $\psi \circ \xi = 1_{H \otimes \hM}$ is more involved, making use of $\nu$ being idempotent (\Cref{lemma:hopf_idempotency}) as well as $\iota$ being an equaliser (\Cref{lemma:hopf_split_eq}).
	This is given in \Cref{fig:hopf_iso2}.
\end{proof}

\begin{figure}[t]
	\centering
	\input{figures/hopf_antipode_lemma}
	\caption{The proof of \Cref{lemma:hopf_mulantipode}.}%
	\label{fig:hopfantipodelemma}
\end{figure}


\end{document}